\pdfoutput=1

\documentclass[iop, apj]{emulateapj}
\usepackage{xspace}
\usepackage{amsmath}
\usepackage{framed} 
\usepackage{txfonts}
\usepackage{epstopdf} 
\usepackage{color}
\usepackage{rotating}
\usepackage{natbib}
\usepackage{ulem}

\def\ave#1{\left\langle #1 \right\rangle}
\newcommand{\simgt}{\lower.5ex\hbox{$\; \buildrel > \over \sim \;$}}
\newcommand{\simlt}{\lower.5ex\hbox{$\; \buildrel < \over \sim \;$}}

\slugcomment{To be submitted to PASJ}

\newdimen\hssize
\def\gcm3{\mathrm{g} / \mathrm{cm}^3}

\def\gtsima{$\; \buildrel > \over \sim \;$}
\def\ltsima{$\; \buildrel < \over \sim \;$}
\def\prosima{$\; \buildrel \propto \over \sim \;$}
\def\gsim{\lower.7ex\hbox{\gtsima}}
\def\lsim{\lower.7ex\hbox{\ltsima}}
\def\simgt{\lower.7ex\hbox{\gtsima}}
\def\simlt{\lower.7ex\hbox{\ltsima}}
\def\simpr{\lower.7ex\hbox{\prosima}}

\shorttitle{Detection of Universal DM Profile with Subaru Weak Lensing}
\shortauthors{Niikura et al.}

\begin{document}

%
\def\figdir{.}
\def\figext{pdf}


\title{Detection of universality of dark matter profile from Subaru
weak lensing measurements of 50 massive clusters}
\author{Hiroko Niikura \altaffilmark{1,2},
Masahiro Takada\altaffilmark{1},
Nobuhiro Okabe\altaffilmark{1,3},
Rossella Martino\altaffilmark{4},
Ryuichi Takahashi\altaffilmark{5}
}

\affil{
$^1$ Kavli Institute for the Physics and Mathematics of the Universe
(Kavli IPMU, WPI), UTIAS,
The
University of Tokyo, 
Chiba, 277-8583, Japan \\
$^2$ Physics Department, The University of Tokyo, Bunkyo, Tokyo
113-0031, Japan\\
$^3$Department of Physical Science, Hiroshima University, 1-3-1 Kagamiyama, Higashi-Hiroshima, Hiroshima 739-8526, Japan\\
$^4$ Laboratoire AIM, IRFU/Service d'Astrophysique-CEA-CNRS,
Bt. 709, CEA-Saclay, 91191 Gif-sur-Yvette Cedex, France \\
$^5$Faculty of Science and Technology, Hirosaki University, 3 Bunkyo-cho, Hirosaki, Aomori 036-8561, Japan
}
   \begin{abstract}
   We develop a novel method of measuring the lensing distortion
   profiles of clusters with stacking the ``scaled'' amplitudes of
   background galaxy ellipticities as a function of the ``scaled''
   centric radius according to the Navarro-Frenk-White (NFW) prediction
   of each cluster, based on the assumption that the different clusters
   in a sample follow the {\it universal} NFW profile.  First we
   demonstrate the feasibility of this method using both the analytical
   NFW model and simulated halos in a suite of high-resolution $N$-body
   simulations. We then apply, as a proof of concept, this method to the
   Subaru weak lensing data and the {\it XMM}/{\it Chandra} X-ray
   observables for a sample of 50 massive clusters in the redshift range
   $0.15\le z\le 0.3$, where their halo masses differ from each other by
   up to a factor of 10.  To estimate the NFW parameters of each
   cluster, we use the halo mass proxy relation of X-ray observables,
   based on either the hydrostatic equilibrium or the gas mass, and then
   infer the halo concentration from the model scaling relation of halo
   concentration with halo mass.  We evaluate a performance of the NFW
   scaling analysis by measuring the scatters of 50 cluster lensing
   profiles relative to the NFW predictions over a range of radii,
   $0.14\le R/[h^{-1}{\rm Mpc}]\le 2.8$. We found a 4 -- 6$\sigma$ level
   evidence of the universal NFW profile in 50 clusters, for both the
   X-ray halo mass proxy relations, although the gas mass appears to be
   a better proxy of the underlying true mass.  By comparing the
   measurements with the simulations of cluster lensing profiles taking
   into account the statistical errors of intrinsic galaxy shapes in the
   Subaru data, we argue that additional halo mass errors or intrinsic
   scatters of $\sigma(M_{500c})/M_{500c}\sim 0.2$ -- $0.3$ could
   reconcile a difference between the measurements and the simulations.
    This method allows us to some extent to preserve characteristics of
   individual clusters in the statistical weak lensing analysis, thereby
   yielding a new means of exploiting the underlying genuine form of the
   halo mass profile and the halo mass proxy relations via weak lensing
   information, under the assumption of the existence of the universal
   profile.
   \end{abstract}


\section{Introduction}

Clusters of galaxies are the largest, gravitationally bound objects in
the Universe, and the formation and evolution processes are dominated by
gravitational effects mainly due to dark matter. Hence clusters provide
us with a useful laboratory for studying the nature of dark matter
\citep{Cloweetal:06} as well as constraining cosmology, e.g. from the
abundance of clusters found from a survey volume
\citep{Vikhlininetal:09,Rozoetal:10,OguriTakada:11}. However, to attain
the full potential of cluster-based cosmology from upcoming wide-area
surveys such as the Subaru Hyper
Suprime-Cam\footnote{\url{http://www.naoj.org/Projects/HSC/j\_index.html}}
and the Dark Energy
Survey\footnote{\url{http://www.darkenergysurvey.org}} requires a
further understanding of the physical processes in clusters.

One of the most important predictions in $N$-body simulations of the
$\Lambda$-dominated, cold dark matter structure formation model
($\Lambda$CDM) is the emergence of universal mass density profile --
that is, the mass density profile of dark matter halos can be well
fitted by a ``universal'' two-parameter family of the model profile over
a wide range of halo masses, first proposed in \citet[][hereafter
NFW]{NFW:96,NFW:97}.  The NFW profile predicts a monotonically steepened
profile with increasing radius, with logarithmic slopes shallower than
an isothermal sphere interior to the characteristic ``scale'' radius
$r<r_s$, but steeper at larger radius, approaching $r^{-3}$ at the
virial radius, $r\rightarrow r_{\rm vir}$ \citep[see also][for
discussion on the physical origin within the framework of the
hierarchical $\Lambda$CDM model]{Dalaletal:10}.  Further, the ratio of
the characteristic scale radius to the virial radius, which
characterizes the degrees of central concentration of the mass
distribution -- the so-called halo concentration $c$ -- tends to be lower
for more massive halos. In addition the halo concentration of a fixed
halo mass displays intrinsic scatters typically given by $\sigma_{\ln
c}\sim 0.2$, originating from details of the mass accretion or assembly
history of each halo in the hierarchical structure formation
\citep{Bullocketal:01,Wechsleretal:02,Zhaoetal:03,Duffyetal:08,Zhaoetal:09,Bhattacharyaetal:13,DiemerKravtsov:14}.
Thus these properties of dark matter halos are important predictions of
$\Lambda$CDM model, and need to be carefully tested by measurements.

Gravitational lensing is a unique, powerful method enabling one to probe
the matter distribution in galaxy clusters irrespective of their
physical and dynamical states \citep{Schneider:06}. Several works have
investigated the mass density profile over a wide range of radii by
combining the strong and weak lensing at the small and large radii,
observationally the exquisite high-resolution images of Hubble Space
Telescope and the wide-field ground based telescopes such as the Subaru
Telescope
\citep{TysonFischer:95,Kneibetal:03,Broadhurstetal:05,Smithetal:05,Ogurietal:05,Ogurietal:12,Hoekstraetal:12,Newmanetal:13,Zitrinetal:14,Mertenetal:14}.
In addition, the stacked weak lensing analysis combining a sample of
clusters has been proven to be a robust, powerful method of probing the
average mass distribution of the sampled clusters
\citep{Johnstonetal:07,Okabeetal:10,Ogurietal:12,Okabeetal:13,Umetsuetal:14}.
These works have shown that the average mass profile measured from the
stacked lensing is in remarkably nice agreement with the NFW prediction.
Another advantage of stacked lensing is it allows one to probe the
mass distribution even for less massive halos, such as galaxy-scale
halos, as long as a sufficient number of sampled halos (e.g. galaxies)
are used in the analysis
\citep{Mandelbaumetal:05,Leauthaudetal:10,Miyatakeetal:13}. However, a
downside of the stacked lensing method is the loss of lensing
information of individual clusters. Hence a knowledge of the
distribution of the underlying halo parameters in the sampled clusters
such as their halo masses is of critical importance in order not to have
any bias in the NFW parameters inferred from the stacked lensing signals
\citep{OguriTakada:11}. This is equivalent to the importance of
exploring a well-calibrated proxy relation of cluster observables
 with halo mass or more generally halo parameters 
\citep{Rozoetal:09,Zhangetal:10,Okabeetal:10b,Zhangetal:11,Mahdavietal:13,vanderLindenetal:14,Martinoetal:14,Okabeetal:14,Hoekstraetal:15}.

The purpose of this paper is to develop a novel method of measuring the
lensing distortion profiles of clusters, motivated by the NFW
prediction. We propose the ``NFW scaling'' analysis for the lensing
measurements, which is done by averaging the ``scaled'' amplitudes of
background galaxy ellipticities in each bin of the ``scaled'' radii
according to the NFW prediction of individual cluster.  With this NFW
scaling method, we can address whether clusters display
the universality of their lensing profiles as seen in simulations.
First, to demonstrate the feasibility of the NFW scaling analysis, we
will use simulations of cluster lensing observables based on a suite of
high-resolution $N$-body simulations. Then, as a proof of concept, we
will apply this method to a sample of 50 massive clusters in the
redshift range $0.15\le z\le 0.3$ that have been observed with the
Subaru telescope by the Local Cluster Substructure Survey \citep[LoCuSS,
and also see][for
details]{Okabeetal:10,Okabeetal:13,Martinoetal:14}\footnote{Based in
part on data collected at Subaru Telescope and obtained from the SMOKA,
which is operated by the Astronomy Data Center, National Astronomical
Observatory of Japan.}. Note that this study is based on the published
results of LoCuSS, and is not performed within the collaboration.
To estimate the NFW scaling of each cluster, we will use the halo mass
estimate in \citet{Martinoetal:14} based on the {\it XMM} and/or {\it
Chandra} X-ray observables, and use the halo concentration inferred from
the model scaling relation between halo mass and concentration in
\citet{DiemerKravtsov:14}.  Then by comparing the scatters of 50 cluster
lensing profiles relative to the NFW predictions for two cases with and
without the NFW scaling, we test the performance of this method as well
as the universality of the cluster mass distribution.

The structure of this paper is as follows. In \S~\ref{sec:method}, after
briefly reviewing the lensing observables of NFW halo, we will derive an
estimator of the lensing distortion profile measurement with NFW
scaling. Then we study the feasibility of this method using analytical
NFW models and $N$-body simulations. In \S~\ref{sec:results}, we first
describe the Subaru weak lensing catalog and the X-ray observables for
the sample of massive clusters we use in this paper, and show the
results of the application of this method to the Subaru data.
\S~\ref{sec:conclusion} is devoted to discussion and conclusion.  Unless
stated otherwise, we will adopt a flat $\Lambda$CDM cosmology with
$\Omega_m=0.27$, $\Omega_\Lambda=0.73$, and the Hubble parameter
$h=H_0/(100~{\rm km~s}^{-1}{\rm Mpc}^{-1})=0.70$.

\section{Methodology: Stacked weak lensing with NFW scaling}
\label{sec:method}

\subsection{Lensing of Navarro-White-Frenk halo}

The Navarro-Frenk-White (1997; hereafter NFW) mass density profile for a
halo is parametrized by two parameters as
\begin{equation}
\rho_\mathrm{NFW}(r)= \frac{\rho_{c}}{(r/r_{s})(1+r/r_{s})^2},
\label{eq:rho_nfw}
\end{equation}
where $r_s$ is the scale radius and $\rho_c$ is the central density
parameter. The parameter $\rho_c$ is specified by imposing that the mass
enclosed within a sphere of a given overdensity $\Delta$ is equal to the
halo mass $M_\Delta$,
\begin{equation}
 \rho_c=\frac{\Delta \rho_{\rm cr}(z)c_\Delta^3}{3m_{\rm NFW}(c_\Delta)}
  =\frac{M_\Delta}{4\pi r_s^3m_{\rm NFW}(c_\Delta)},
  \label{eq:rhoc}
\end{equation}
where $m_{\rm NFW}(c_\Delta)\equiv \int^{c_\Delta}_0dx~ x/(1+x)^2
=\ln(1+c_\Delta)- c_\Delta/(1+c_\Delta)$, $c_\Delta \equiv
r_\Delta/r_s$, a concentration parameter, and $\Delta(z)$ is a nonlinear
overdensity introduced to define the interior mass for each halo.  Note
that throughout this paper we employ halo mass definition with respect
to the critical density, not the mean mass density: $M_\Delta\equiv
(4\pi/3)r_\Delta^3 \rho_{\rm cr}(z)\Delta$.
Although we focus on the NFW profile in this paper, the
method developed in this paper can be applied to any other univeral
profile such as generalized NFW or Einasto profile \citep{Einasto:65,Merrittetal:06}.

Several works have shown a scaling relation of the halo concentration
with halo mass, using numerical simulations or based on analytical
arguments
\citep{Bullocketal:01,Wechsleretal:02,Zhaoetal:03,Duffyetal:08}
\citep[most recently][hereafter DK15 and see references
therein]{DiemerKravtsov:14}. As for our fiducial model, we adopt the
publicly-available code provided by B. Diemer to compute the halo mass
and concentration relation in DK15.  Note that we used the ``median''
relation, rather than the mean, for our default choice as recommended in
DK15.  The mass estimates from the X-ray observables are not $M_{200c}$,
and rather the interior mass of a greater overdensity such as
$M_{500c}$. Assuming that a halo exactly follows the NFW profile, we can
convert the scaling relation calibrated for $M_{200c}$ to the
$c_{500c}$-$M_{500c}$ relation, based on the method in
\citet{HuKravtsov:03}. The public code of DK15 allows us to compute the
halo concentration for an input overdensity based on this method. We
will also study how possible variations in the $c$-$M$ relation affect
the results of this paper.

For an NFW profile, we can derive an analytical expression for the
lensing convergence and shear profiles
\citep{Bartelmann:96,GolseKneib:02}:
\begin{eqnarray}
 \kappa^{\rm NFW}(R)&\equiv & \frac{\Sigma^{\rm NFW}(R)}{\Sigma_{\rm
  cr}(z_\mathrm{l},z_\mathrm{s})}=2\rho_cr_s\frac{g^{\rm NFW}(R/r_s)}{\Sigma_{\rm crit}(z_\mathrm{l},z_\mathrm{s})},\nonumber\\
\gamma^{\rm NFW}_+(R)&\equiv & \frac{\Delta\Sigma^{\rm NFW}(R)}{\Sigma_{\rm
  cr}(z_\mathrm{l},z_\mathrm{s})}=2\rho_cr_s\frac{f^{\rm NFW}(R/r_s)}{\Sigma_{\rm
  crit}(z_\mathrm{l},z_\mathrm{s})},
\label{eq:nfw_lens}
\end{eqnarray}
where $R$ is the projected comoving radius from halo center, and the
functions $f^{\rm NFW}(x)$ and $g^{\rm NFW}(x)$ are given by
\begin{equation}
 g^{\rm NFW}(x)
  =\left\{
  \begin{array}{ll}
  {\displaystyle \frac{1}{x^2-1}\left(1-\frac{1}{\sqrt{1-x^2}}\mathrm{cosh}^{-1} \frac{1}{x}\right)}, & (x<1)\\
  {\displaystyle \frac{1}{3}}, & (x=1)\\
  {\displaystyle \frac{1}{x^2-1}\left(1-\frac{1}{\sqrt{x^2-1}}\mathrm{cos}^{-1} \frac{1}{x}\right)}, &(x>1)\\
  \end{array}
  \right.
 \end{equation}
 and
 \begin{eqnarray}
&& f^{\rm NFW}(x)\nonumber\\
&&  =\left\{
  \begin{array}{ll}
   {\displaystyle \frac{2}{x^2}\ln\frac{x}{2}+\frac{1}{1-x^2}
    \left(
1+\frac{2-3x^2}{x^2\sqrt{1-x^2}}\mathrm{cosh}^{-1}\frac{1}{x}
	     \right)},
    & (x<1)\\
   {\displaystyle \frac{5}{3}-2\ln2}, &  (x=1)\\
      {\displaystyle \frac{2}{x^2}\ln\frac{x}{2}-\frac{1}{x^2-1}
    \left(
1+\frac{2-3x^2}{x^2\sqrt{x^2-1}}\mathrm{cos}^{-1}\frac{1}{x}
	     \right)},
 & (x>1).\\
  \end{array}
			 \right.
  \label{eq:fnfw}
 \end{eqnarray}
The critical surface mass density $\Sigma_{\rm crit}$ for a given system
of lens cluster and source at redshifts $z_{\mathrm{l}}$ and
$z_{\mathrm{s}}$, respectively, is given as
\begin{equation}
 \Sigma_{\rm crit}(z_\mathrm{l},z_\mathrm{s})
  = \frac{c^2}{4\pi
  G}\frac{D_\mathrm{A}(z_\mathrm{s})}{D_\mathrm{A}(z_\mathrm{l})D_\mathrm{A}(z_\mathrm{l},z_\mathrm{s})(1+z_{\rm
  l})^2},
\end{equation}
where $D_A(z)$ is the angular diameter distance and the factor $(1+z_{\rm
l})^2$ is from our use of the comoving scale.  From Eqs.~(\ref{eq:rhoc})
and (\ref{eq:nfw_lens}), we can find that the lensing amplitudes of an
NFW halo scale with the NFW parameters ($M_\Delta, c_\Delta$) as
\begin{eqnarray}
 \kappa^{\rm NFW}, \gamma_+^{\rm NFW}&\propto& 2\rho_c r_s\propto
  M_{\Delta}\left/\left(r_s^2m_{\rm NFW}(c_\Delta)\right)\right.\nonumber\\
  &\propto& 
  M_\Delta^{1/3} c_{\Delta}^2/m_{\rm NFW}(c_\Delta).
  \label{eq:nfwlens_cM}
\end{eqnarray}
If we employ the $c_\Delta$-$M_\Delta$ scaling relation given as
$c_\Delta(M_\Delta)\propto M^{-\alpha}$, the lensing amplitudes roughly
scale with halo mass as $\gamma_+^{\rm NFW}\propto M^{1/3-2\alpha}$ as
the function $m(c_{\Delta })$ has a weak dependence on halo mass.  Note
that, since the cluster sample is among the most massive
clusters, we have checked that the 2-halo term is much smaller than the
above 1-halo term, by a factor of 100, over a range of the radii we
consider \citep[e.g., see][]{OguriTakada:11,TakadaSpergel:14}. Therefore we ignore the
2-halo term for the following analysis.

An actual lensing observable estimated from ellipticities of
background galaxies for an NFW lens is the lensing ``distortion'' profile
or reduced shear profile:
\begin{equation}
 \ave{e_+}(R)\rightarrow 
  \frac{\gamma_+^{\rm NFW}(R)}{1-\kappa^{\rm NFW}(R)},
\label{eq:reduced_shear}
\end{equation}
where $e_+$ is the tangential component of the ellipticities with
respect to cluster center. The reduced shear correction is not
negligible at the inner radii, and we need to take into account the
correction.

\subsection{Stacked lensing without NFW scaling}

For the standard method to estimate the stacked lensing profile for
$N_c$ clusters,  we follow the method in
\citet{Johnstonetal:07} and \citet{Mandelbaumetal:13}:
  \begin{equation}
    \widehat{\ave{\Delta\Sigma}}(R)=
    \frac{1}{N}\sum_{a=1}^{N_c}\sum_{s_a; |{\bf R}_{(a)s_a}|\in R}
    w_{(a, s_a)}\Sigma_{{\rm cr}(a)}
    e_{(s_a)+}({\bf R}_{s_a}),
    \label{eq:est_dSigma}
   \end{equation}
where $e_{(s_a)+}$ is the tangential ellipticity of the $s_a$-th
background galaxy in the $a$-th cluster region, and $N$ is the
normalization factor defined as
\begin{equation}
 N=\sum_{a=1}^{N_c}\sum_{s_a} w_{(a,s_a)}.
\end{equation}
The summation $\sum_{a}$ runs over the sampled clusters, from $a=1$ to
$N_c $,
and the summation $\sum_{s_a; |{\bf R}_{s_a}|\in
R}$ runs over all the background galaxies that reside in the annulus of
radius $R$ from the $a$-th cluster center to within the bin width.  We
employ the weight given as
\begin{equation}
w_{(a,s_a)}=\frac{1}{\Sigma_{\rm
cr}(z_a,z_{s_a})^2(e_{(s_a)}^2+\sigma_{(s_a)e}^2+\alpha^2)},
\label{eq:w}
\end{equation}
where $z_{s_a}$ is the redshift of the $s_a$-th background galaxy,
$e_{(s_a)}$ is the ellipticity amplitude, $\sigma_{(s_a)e}$ is the
measurement error and $\alpha$ is the constant factor to regularize the
weight for which we adopt $\alpha=0.4$ \citep{Okabeetal:10}. Note that
we employ the average redshift for all the source galaxies in each
cluster region, as described below in detail.

Since we need to employ a finite number of the radial bins to study the
``shape'' of lensing distortion profile, which binning scheme to use is
not so clear.  As for the representative value of a given radial bin, we
use the average of radii of background galaxies that reside in the
annulus taking into account their weights:
\begin{equation}
 R \equiv \frac{\sum_{a=1}^{N_c}\sum_{s_a; |{\bf R}_{(a)s_a}|\in
  R}w_{(a, s_a)}R_{(a)s_a}}{\sum_{a=1}^{N_c}\sum_{s_a; |{\bf
  R}_{(a)s_a}|\in R}w_{(a, s_a)} }.
  \label{eq:r}
\end{equation}
In the literature the area-weighted value of each radial bin is often
used. We have checked that, using an analytical NFW profile and taking
the actual distribution of background galaxies in the Subaru data, the
above radial binning is more accurate in the sense that the distortion
profile is in better agreement with the model NFW profile amplitude
inferred by the representative value of the radial bin, less than 1\% in
the fractional difference for most cases.

The statistical uncertainty of the stacked lensing at each radial bin
can be estimated as
\begin{equation}
%
   \sigma_{\ave{\Delta \Sigma}}(R)^2=\frac{1}{2N^2}
   \sum_{a=1}^{N_c}\sum_{s_a; |{\bf R}_{(a)s_a}|\in R} w_{(a,s_a)}^2\Sigma_{{\rm cr}(a)}(z_a,z_{s_a})^2e_{(s_a)}^2.
\label{eq:e_est_dSigma}
\end{equation}
In this paper we consider the intrinsic ellipticities as a source of the
statistical errors in the lensing measurement, and ignore the cosmic
shear contribution that arises from different mass distribution along
the same line of sight to the cluster. For the application of this
method to the Subaru data that we will show below,
this is 
a good approximation in practice, because the mean number density of background
galaxies is small, about $5$ arcmin$^{-2}$, after a secure selection of
background ``red'' galaxies as we will discuss in \S~\ref{sec:subaru}
\citep[also see][for the details]{Okabeetal:13}.

When comparing the measured lensing profile to an NFW model, we need to
account for the contribution of reduce shear.  In this paper, assuming
that all the clusters follow a single NFW profile in average sense, we
model the the stacked lensing profile, according to
Eqs.~(\ref{eq:nfw_lens}) and (\ref{eq:reduced_shear}) as
\begin{eqnarray}
 \ave{\widehat{\Delta\Sigma}}(R)&\iff &\frac{\Delta\Sigma^{\rm
  NFW}\left(R\right)}{1-\kappa^{\rm NFW}\left(R\right)}\nonumber\\
  &\simeq& \Delta\Sigma^{\rm
  NFW}\left(R\right)\left[
       1+\ave{\frac{1}{\Sigma_{\rm cr}}}\Sigma^{\rm NFW}\left(\ave{R}\right)
		  \right],
  \label{eq:model_redshear}
\end{eqnarray}
where the notation ``$\iff$'' is meant to denote the comparison between
the measurement (left-hand side) and the model profile (right-hand
side).  The notation $\ave{\hspace{1em}}$ on the right-hand side denotes
the average taking into account the weights of background galaxies in
each cluster region as in Eq.~(\ref{eq:r}).  We will use the above
equation to estimate the halo mass and concentration parameter,
$M_\Delta$ and $c_\Delta$, from the measured lensing profile.

\subsection{Stacked lensing with NFW scaling}
\label{sec:stacked_nfw}

Now we consider the stacked lensing analysis with ``NFW scaling''.  To
implement this method we combine the weak lensing measurement and X-ray
observables, where the X-ray observables are needed to estimate halo
mass of each cluster independently of the lensing observables.  Assuming
that each of the sampled clusters follows an NFW profile specified by
their respective parameters, $M_{(a)}$ and $c_{(a)}$, we can define an
estimator of the {\it normalized} NFW lensing profile from the measured
ellipticities of background galaxies, as motivated by
Eq.~(\ref{eq:nfw_lens}):
\begin{equation}
    \widehat{\ave{f^{\rm NFW}}}(x)=
    \frac{1}{N}\sum_{a=1}^{N_c}\sum_{s_a; |{\bf x}_{(a)s_a}|\in x}
    \frac{    w_{(a, s_a)}\Sigma_{{\rm cr}(a)}
    e_{(s_a)+}({\bf x}_{s_a})}{2\rho_{c}\!\left(M^{\rm X}_{(a)},c^{\rm
    X}_{(a)}\right)r_s\!\left(M^{\rm X}_{(a)},c^{\rm X}_{(a)}\right)}.
    \label{eq:est_fnfw}
\end{equation}
Here $M^{\rm X}_{(a)}$ and $c^{\rm X}_{(a)}$ are the halo mass and
concentration for the $a$-th cluster, estimated from the X-ray
observables (see below for details).  The {\it scaled} radius in the
above equation, $x$, is defined for the $a$-cluster as ${\bf
x}_{(a)s_a}\equiv {\bf R}_{(a)s_a}/r_{s}(M^{\rm X}_{(a)},c^{\rm
X}_{(a)})$, where $r_s$ is the scale radius of NFW profile,
$r_{s}=r_\Delta/c_\Delta$.  We use the representative value of each
radial bin, $x$, defined in a similar manner to Eq.~(\ref{eq:r}).  The
central density parameter of NFW profile, $\rho_c$, can be estimated
from $M^{\rm X}_{(a)}$ and $c^{\rm X}_{(a)}$ for the $a$-th cluster,
from Eq.~(\ref{eq:rhoc}).  Note that the profile $\widehat{\ave{f^{\rm
NFW}}}$ and the radius $x$ are dimension-less. With the above NFW
scaling, weak lensing signals due to less massive halos than the mean
mass in the sampled clusters are up-weighted, while the signals of more
massive halos are down-weighted.

Similarly, the measurement errors of the stacked profile at each radial
bin are estimated as
\begin{equation}
\sigma_{\ave{f^{\rm NFW}}}(x)^2=\frac{1}{2N^2}
   \sum_{a=1}^{N_c}\sum_{s_a; |{\bf x}_{(a)s_a}|\in x} \frac{w_{(a,s_a)}^2\Sigma_{{\rm
   cr}(a)}(z_a,z_{s_a})^2e_{(s_a)}^2}
   {4\rho_{c}\!\left(M^{\rm X}_{(a)},c^{\rm X}_{(a)}\right)^2
   r_s\!\left(M^X_{(a)},c^X_{(a)}\right)^2}.
\label{eq:e_est_fnfw}
\end{equation}

To test an improvement in the stacked lensing analysis of NFW scaling
compared to the standard stacked lensing, we compare the scatters of
lensing distortion profiles of the sampled clusters relative to the NFW
prediction. To be more precise, we quantify the scatters by
\begin{equation}
 d^2\equiv \sum_{a=1}^{N_c}\sum_{i}
\frac{\left[
\widehat{\Delta\Sigma_{(a)}}(R_{(a)i})
-\Delta\Sigma^{\rm bf-NFW}\!\left(R_{(a)i};M_{\rm bf}, c_{\rm bf}\right)\right]^2}
{\sigma_{\Delta\Sigma(a)}(R_{(a)i})^2}
\label{eq:d}
\end{equation}
or
\begin{equation}
 d^2_{\rm w-scaling}\equiv \sum_{a=1}^{N_c}\sum_{i}
\frac{\left[
\widehat{f^{\rm NFW}_{(a)}}\!\left(x_{(a)i}\right)-f^{\rm
NFW}\!\left(x_{(a)i}\right)\right]^2}{\sigma_{f^{\rm NFW}(a)}(x_{(a)i})^2}.
\label{eq:d_nfw}
\end{equation}
Here $\widehat{\Delta\Sigma_{(a)}}$ and $\widehat{f^{\rm NFW}_{(a)}}$
are the measured distortion profile without and with NFW scaling for the
$a$-th cluster, which are estimated in the similar manner to
Eqs.~(\ref{eq:est_dSigma}) and (\ref{eq:est_fnfw}), and $\sigma_{\Delta
\Sigma (a)}$ and $\sigma_{f^{\rm NFW}(a)}$ are the errors at each radial
bin, estimated similarly to Eqs.~(\ref{eq:e_est_dSigma}) and
(\ref{eq:e_est_fnfw}), respectively.  $\Delta\Sigma^{\rm bf-NFW}(R)$ is
the best-fit NFW profile of the stacked lensing profile
(Eq.~\ref{eq:model_redshear}). For the NFW scaling case, we similarly
include the reduced shear correction: we multiply the function $f^{\rm
NFW}(x)$ (Eq.~\ref{eq:fnfw}) by the function, $1+\ave{1/\Sigma_{\rm
cr}}_{w_{(a,s_a)}}\Sigma^{\rm NFW}(x)$ as in
Eq.~(\ref{eq:model_redshear}), where we used the best-fit NFW model of
the stacked lensing profile without NFW scaling in order to compute
$\Sigma^{\rm NFW}(x)$. The above $d^2$ and $d^2_{\rm w-scaling}$ are
equivalent to the log-likelihood functions of lensing distortion
profiles of $N_c$ clusters assuming that the statistical errors are
given by the intrinsic ellipticities. The radial bin $R_{(a)i}$ or
$x_{(a)i}$ for the $a$-th cluster is similarly computed by
Eq.~(\ref{eq:r}) from the background galaxies that reside in the annulus
of the cluster. However, comparing the scatters of lensing distortion
profiles with and without NFW scaling requires a careful treatment of
the radial binning. As we will later describe in more detail, we will
employ the radial binning scheme so as to preserve the same background
galaxies in the $i$-th radial bins with or without the NFW scaling.
With this binning scheme, the relation $d^2=d^2_{\rm w-scaling}$ holds
if setting the model profiles to $\Delta\Sigma^{\rm bf-NFW}=f^{\rm
NFW}=0$ \footnote{If we set $\Delta\Sigma^{\rm bf-NFW}=f^{\rm NFW}=0$ in
Eqs.~(\ref{eq:d}) and (\ref{eq:d_nfw}), the values of $d^2$ give the
cumulative signal-to-noise ratio of the lensing distortion measurements
for the $N_c$ clusters.}.  If the lensing distortion profiles of the
sampled clusters are similar in their shapes and amplitudes, following
the NFW profile, the value of $d^2$ should become smaller: $d^2_{\rm
w-scaling}<d^2$.

\subsection{Testing the method with $N$-body simulations}
\label{sec:nbody}
 \begin{figure}
  \centering{
\includegraphics[scale=1.0]{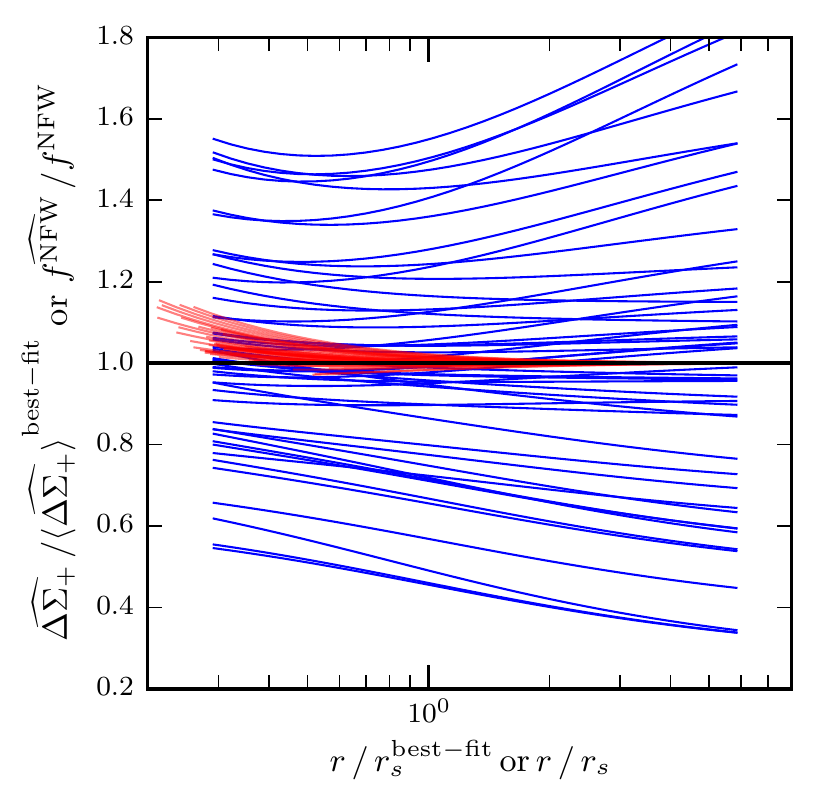}}
  \caption{The distribution of NFW lensing profiles for 50 halos for
  each of which we took the X-ray inferred mass of 50 Subaru clusters
  (here the hydrostatic equilibrium mass in Table~\ref{tab:x-ray}) and
  assumed the halo concentration based on the halo mass and
  concentration relation, $c=c(M_{500c})$, in \citet[][hereafter
  DK15]{DiemerKravtsov:14}. The blue curves are the lensing profiles
  without ``NFW scaling'', i.e. the standard method, but each curve is
  normalized by the best-fit NFW profile to the stacked profile of 50
  profiles, and is plotted as a function of the radius relative to the
  scale radius of the best-fit NFW model. The red curves are the lensing
  profiles with ``NFW scaling'', computed assuming that the halo mass
  and concentration of each halo are {\it a priori} known -- an ideal
  case. Note that we fixed the same range of radii, $0.14<R/[h^{-1}{\rm
  Mpc}]<2.8$, in the comoving length units for both the results. }
  \label{fig:test_nfw}
 \end{figure}

In this subsection, before going to the Subaru data, we test our method
using analytical NFW model and high-resolution $N$-body simulations.
For the sake of convenience to compare with the following sections, we
consider 50 clusters in this section as the 50 Subaru clusters.

First let's consider an ideal case, albeit unrealistic, that each of 50
clusters {\it exactly} follows an NFW profile. Figure~\ref{fig:test_nfw}
shows the lensing profiles with or without the NFW scaling for 50 halos.
To take into account variations in halo masses that resemble the 50
clusters, we assign one-by-one the X-ray inferred masses of 50 cluster
to NFW halos\footnote{We here employed the hydrostatic equilibrium mass
in \citet{Martinoetal:14}, which was estimated from the X-ray
observables of each cluster.}.  Note that we use the $c$-$M$ scaling
relation in DK15 to compute the halo concentration for each NFW halo.
The different blue curves show each NFW distortion profile relative to
the best-fit NFW model of the stacked distortion profile, as a function
of the radius relative to the scale radius of the best-fit NFW
model. Here we consider the same range of radii, $0.14\le R/[h^{-1}{\rm
Mpc}]\le 2.8$ for all the halos as we will do for actual analysis of
Subaru data. For the range of cluster masses, the lensing distortion
amplitudes differ from each other by up to a factor of 5.

On the other hand, the red curves in Figure~\ref{fig:test_nfw} show the
profiles after the NFW scaling implementation, assuming that the true
mass and concentration of each cluster are {\it a priori} known, i.e. an
ideal case. Each curve is the fractional profile relative to the NFW
distortion profile including the reduced shear correction, $f^{\rm
NFW}(x)$ (Eq.~\ref{eq:fnfw} and see below Eq.~\ref{eq:est_fnfw}). The
deviation from unity is due to an imperfect correction of the reduced
shear: the nonlinear correction becomes non-negligible at small radii,
and breaks the universality of the NFW lensing profile. The horizontal
axis is in the units of the ``scaled'' radius, $R/r_s$, where $r_s$ is
the NFW scale radius of each halo. Due to the radial transformation from
the original fixed range of $R$, the range of the scaled radius $x$,
covered by each halo, differ from each other. The figure shows that the
NFW scaling significantly reduces the scatters of lensing profiles,
making the differences within 20\% over a range of radii we consider.

Obviously actual clusters have much more complicated mass distribution
than an analytical NFW model: intrinsic scatters of halo concentration,
aspherical mass distribution, substructures and so on. To study these
effects we use simulated halos of cluster scales, generated from a
high-resolution $N$-body simulation in \citet{Takahashietal:12}.  In
brief the $N$-body simulation was ran with the publicly-available {\it
Gadget-2 code} \citep{Springeletal:01,Springel:05} assuming the WMAP
cosmology.  The simulation employed $1024^3$ particles in a box of
320~$h^{-1}$Mpc on a side. The mass resolution (the particle mass) is
$2.3\times 10^9$~$h^{-1}M_\odot$, so is sufficient to resolve
cluster-scale halos. 

\begin{figure}
  \centering{
  \includegraphics[width=0.44\textwidth]{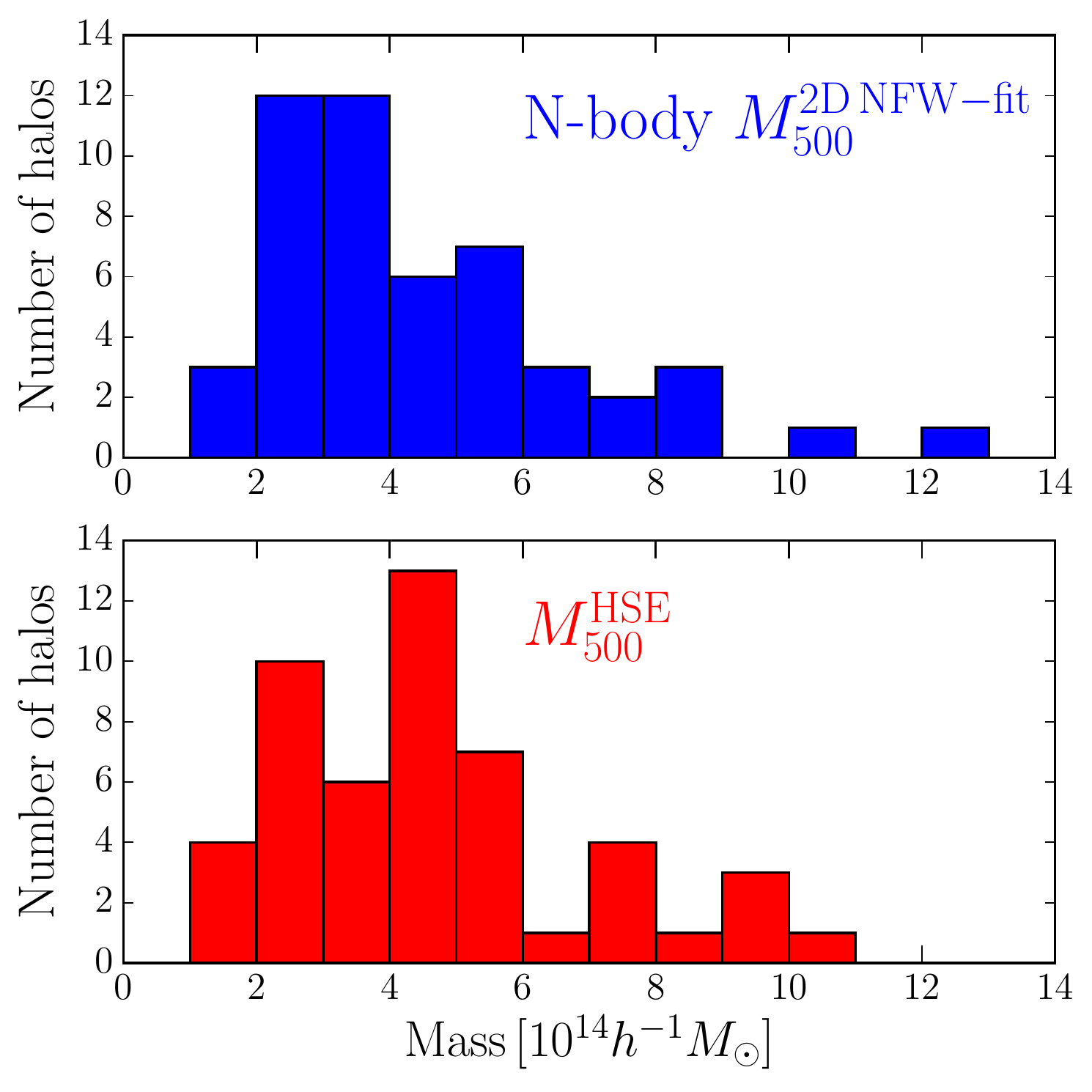}
  \caption{The distribution of halo masses in 50 clusters, taken from
  the $N$-body simulations (see text for details) or estimated based on
  the X-ray observables (here we used the hydrostatic equilibrium
  method). \label{fig:m_nbody_vs_xray} }}
  \end{figure}

To construct a catalog of cluster-scale halos from the $N$-body
simulation output at $z=0$, we used the friends-of-friends (FoF) group
finder \citep[e.g.][]{Davisetal:85} with a linking length of 0.2 in
units of the mean interparticle spacing. For each halo we determined the
halo center using an iterative technique in which the center of mass of
particles within a shrinking sphere is computed recursively until a few
particles are left inside
\citep[e.g.][]{Poweretal:03,Masakietal:13}. Then the halo mass is
defined by a spherical overdensity method -- summing all the particles
within a sphere of a given overdensity $\Delta$ around the halo center.
We constructed a catalog that consists of most massive 50 halos from the
two simulation realizations. Besides the mass threshold, we did not
employ any other selection criteria such as sphericity or the degree of
mass distribution complexity. The mean mass of the selected halos is
similar to the average mass estimated from the lensing measurements of
50 Subaru clusters (see Figure~\ref{fig:m_nbody_vs_xray}).  Exactly
speaking, although the simulated halos are not the {\it same} in detail
as the Subaru clusters, other effects such as the intrinsic
ellipticities of background galaxies cause much larger variations in the
lensing profiles as we will show later. Hence we believe that the
catalog of simulated halos is suitable enough for our purpose.

To test our method as well as to simulate the lensing observables from
the above $N$-body simulations, we use the following procedures:
\begin{itemize}
 \item {\it 3D mass density profile} -- We first computed the
       spherically-average mass profile for each simulated halo,
       $\rho(r)$, where $r$ is the three-dimensional radius from the
       halo center. Then we estimated the NFW parameters, $M_\Delta$ and
       $c_\Delta$ for $\Delta=500$, by fitting the model NFW profile
       (Eq.~\ref{eq:rho_nfw}) to the mass profile, where we weighted the
       simulated mass density profile at a given radial bin by the
       volume of the spherical shell. We stored the best-fit parameters
       ($M^{\rm 3D~ fit}_{500c},c^{\rm 3D~ fit}_{500c}$) for each of the
       50 halos.
 \item {\it 2D lensing profiles} -- To simulate the lensing profiles due
       to a simulated halo, we use the dark matter ($N$-body) particles
       inside or surrounding the halo in the simulation output. We
       estimated the shear profile of each halo by projecting the
       $N$-body particles along the line-of-sight direction:
       \begin{equation}
	\Delta\Sigma(R)=\ave{\Sigma}(<R)-\bar{\Sigma}(R).
       \end{equation}
       Here we chose the $z$-direction of simulation realization for the
       projection, and $R$ is the projected radius from the halo center
       in the $xy$-plane (the plane perpendicular to the projection
       direction). $\ave{\Sigma}(<R)$ is the averaged surface mass
       density within a circle of radius $R$, and $\bar{\Sigma}(R)$ is
       the averaged surface mass density over the annulus of radius $R$.
       In this projection calculation, we used a cubic region containing
       the halo at the center, whose side length is 20~$h^{-1}{\rm
       Mpc}$. Since the shear field arises from the tidal field around a
       halo, the constant mass density field or the mass density field
       beyond the cubic region causes a negligible contribution to
       distortion of background galaxies. We checked that the cubic box
       is large enough for the range of radii we consider. We included
       the reduced shear correction to compute the distortion profile of
       the halo, which is a direct lensing observable:
       \begin{equation}
	\widehat{\Delta
	 \Sigma}(R)=\frac{\Delta\Sigma(R)}{1-\Sigma(R)/\Sigma_{\rm
	 cr}(z_\mathrm{l},z_\mathrm{s})}, 
       \end{equation}
       where $\Sigma_{\rm cr}(z_\mathrm{l},z_\mathrm{s})$ is the lensing
       efficiency.  In doing so we assign the source and cluster
       redshifts of each of the 50 Subaru clusters to each simulated
       halo one-by-one in descending order of halo masses, where we used
       the hydrostatic equilibrium mass of X-ray observables in this
       matching. The assignment of $\Sigma_{\rm cr}$ becomes relevant
       when we will include the effect of background shape noise in the
       Subaru data on the simulated lensing signals of $N$-body
       halos. We estimated the NFW profile parameters, $(M_{500c}^{\rm
       2D},c_{500c}^{\rm 2D})$, by fitting the NFW lensing profile
       (Eq.~\ref{eq:model_redshear}) to the above simulated profile. In
       this fitting we weighted the lensing profile at each radial bin
       by the area of radial annulus.  We stored the distortion profile,
       the lensing efficiency function, $\Sigma_{\rm
       cr}(z_\mathrm{l},z_\mathrm{s})$, and the best-fit NFW parameters
       $(M_{500c}^{\rm 2D~ fit},c_{500c}^{\rm 2D~ fit})$ for each of the
       50 simulated halos.
\end{itemize}

 \begin{figure}
  \centering{
 \includegraphics[width=0.49\textwidth]{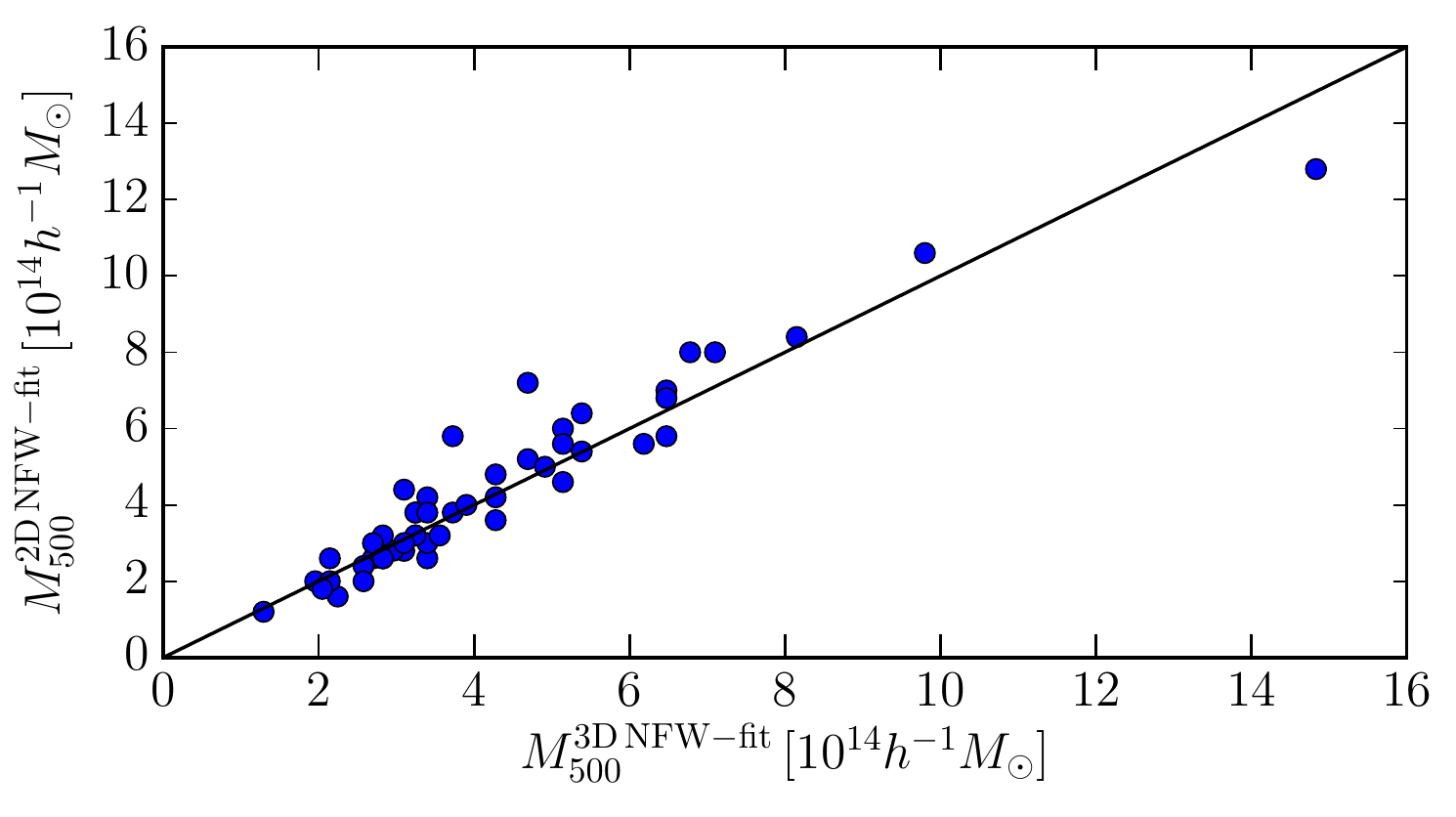}
 \includegraphics[width=0.49\textwidth]{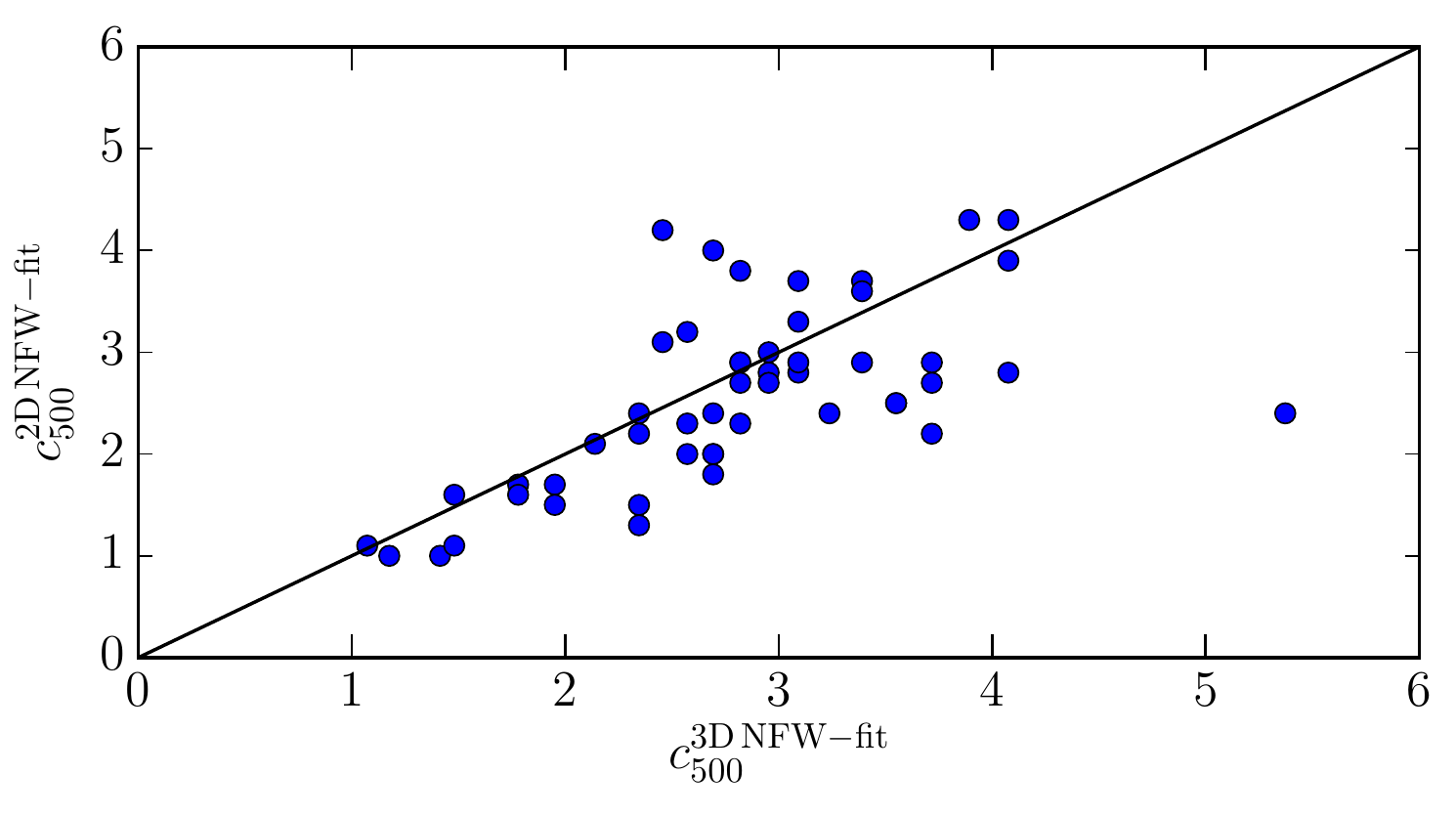}
  \caption{Comparison of the best-fit NFW parameters, halo mass and
  concentration, estimated by fitting the NFW model to the
  three-dimensional mass density profile (``3D NFW-fit'') or the
  two-dimensional lensing distortion profile (``2D NFW-fit''),
  respectively, for the 50 massive halos in $N$-body simulations of
  $\Lambda$CDM model, respectively (see \S~\ref{sec:nbody} for
  details).
  The mean halo mass and the range of their halo masses of simulated
  halos are similar to those of 50 Subaru clusters. \label{fig:m3d_m2d}}}
 \end{figure}
In Figure~\ref{fig:m_nbody_vs_xray} we compare the distribution of halo
masses of simulated halos (upper panel) and 50 clusters (lower), where
we used the best-fit NFW mass for the simulated 2D lensing profile for
each simulated halo and used the X-ray hydrostatic equilibrium mass for
each cluster. The figure shows that the simulated halos cover the
similar range of halo masses as in the Subaru clusters.
 
Figure~\ref{fig:m3d_m2d} compares the best-fit NFW parameters,
$M_{500c}$ and $c_{500c}$, estimated by fitting the NFW model to the
three-dimensional mass density profile or the two-dimensional lensing
distortion profile for each of the 50 simulated halos.  Even if we did
not include any effect of measurement errors, the NFW parameters
inferred from the 3D or 2D fitting generally differ on individual halo
basis. For some halos the 2D fitting halo mass is larger than the 3D
fitting mass, while the 2D concentration is smaller than the 3D
one. These over- or under-estimation would be due to the $c$-$M$
degeneracy in the NFW fitting. These biases might cause a source of
systematic errors in estimating the NFW parameters from the lensing
observables and then testing the $\Lambda$CDM simulation predictions,
e.g. whether or not the $c$-$M$ scaling relation inferred from the
lensing observables is consistent with the $N$-body simulation
predictions \citep[e.g.,][]{Okabeetal:10,Okabeetal:13,Umetsuetal:14}.
This is not the main purpose of this paper, but would be worth to further
study.

 \begin{figure*}[th]
  \centering{
  \includegraphics[width=\textwidth]{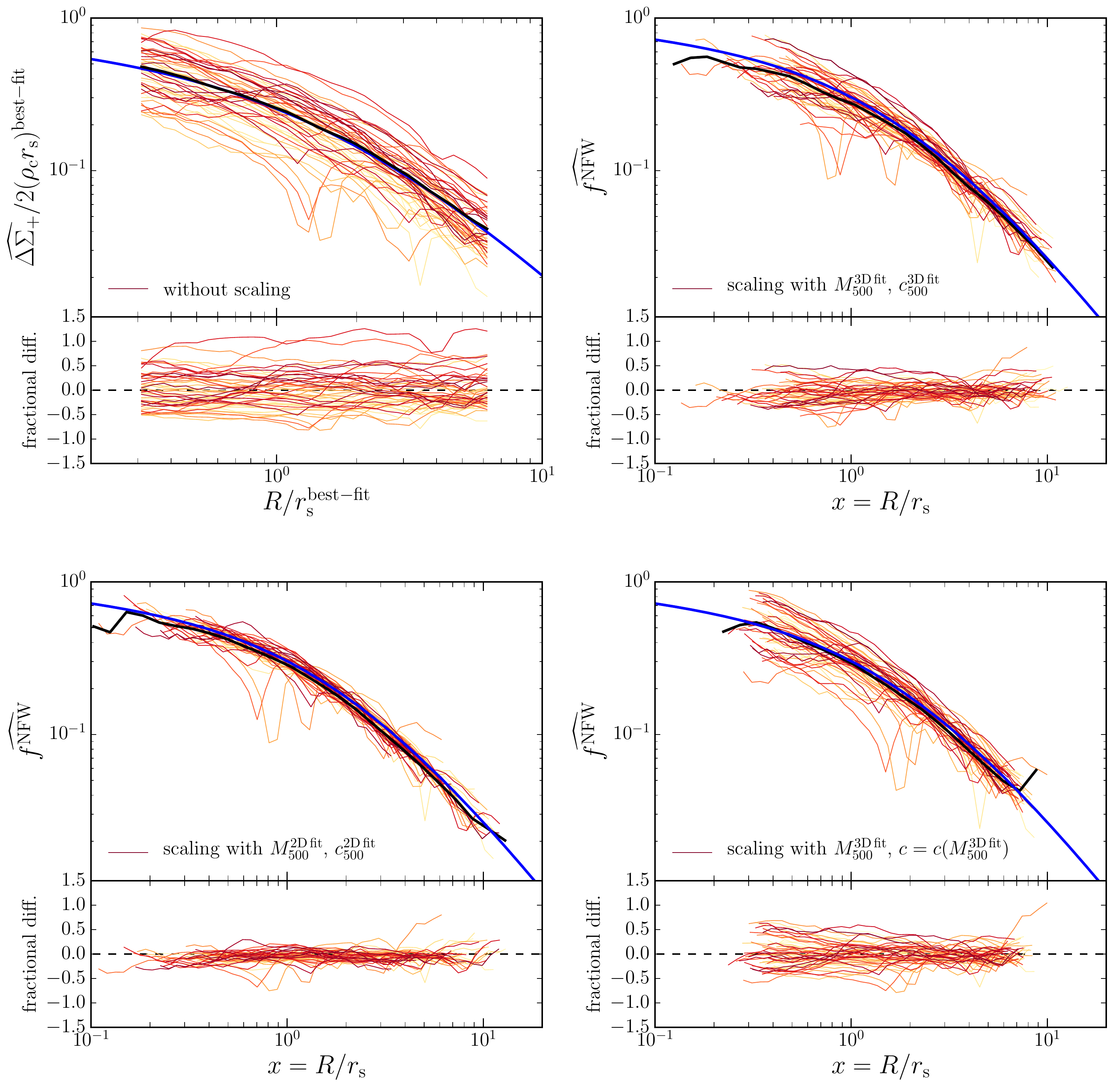}}
\caption{Simulated lensing profiles for 50 massive halos in the $N$-body
  simulation (see \S~\ref{sec:nbody} for details). {\it Upper-left
  panel}: The thin curves show the lensing profiles for each of the 50
  halos, and the bold black curve is the stacked lensing profile without
  NFW scaling. The blue curve is the best-fit NFW profile to the stacked
  profile. As in Figure~\ref{fig:test_nfw}, we plot these profiles in
  terms of the ``scaled'' amplitude,
  $\widehat{\Delta\Sigma_+}/(2\rho_cr_s)^{\rm bf-NFW}$, as a function of
  the ``scaled'' projected radius, $R/r_s^{\rm bf-NFW}$, where we used
  the best-fit NFW parameters of the stacked profile (blue curve). Hence
  both the quantities in the $x$- and $y$-axes are dimension-less, and
  can be directly compared to other panels. The lower plot in each panel
  shows the fractional difference of each profile relative to the
  best-fit NFW profile. The other three panels show the lensing profiles
  for the same halos when implementing the NFW scaling analysis for each
  halo or for the stacked analysis (Eq.~\ref{eq:est_fnfw}). Note that
  the same color curves in the different panels correspond to the same
  halo. {\it Upper-right panel}: The lensing profile when using the NFW
  parameters of three-dimensional mass profile for each halo, $(M^{{\rm
  3D~ NFW-fit}}_{500c}, c^{{\rm 3D~ NFW-fit}}_{500c})$, in the NFW
  scaling analysis. {\it Lower-left panel}: The results when using the
  NFW parameters of two-dimensional lensing distortion profile,
  $(M^{{\rm 2D~ NFW-fit}}_{500c}, c^{{\rm 2D~ NFW-fit}}_{500c})$.  {\it
  Lower-right panel}: Similar to the upper-right panel, but using the
  best-fit halo mass of each halo and using the halo concentration
  inferred from the scaling relation, $c_\Delta=c_\Delta(M_\Delta; z)$
  in DK15. In these three panels, the blue curve is not a fit, but
  rather the NFW prediction itself, $f^{\rm NFW}(x)$
  (Eq.~\ref{eq:fnfw}), where we included the reduced shear correction
  from the best-fit NFW model to the stacked lensing profile in the
  upper-left panel.
  Note that, for all the results, we ignored effects of measurement
  errors such as intrinsic ellipticities of background galaxies.  }
  \label{fig:test_nbody}
 \end{figure*}
In Figure~\ref{fig:test_nbody} we performed a hypothetical experiment of
the stacked lensing analyses with and without NFW scaling, using the 50
simulated halos. Note that we here ignored shape noise contribution for
simplicity. First, the upper-left panel shows the stacked lensing
profile as well as the lensing profiles of individual halos, without NFW
scaling, i.e. based on the standard method. Again note that we used a
fixed range of radial bins, $0.14\le R/[h^{-1}{\rm Mpc}]\le 2.8$ as we
will do for the real data.  The scatters of individual lensing profiles
are significant over a range of the radii.  Each profile shows various
features due to the aspherical mass distribution, in contrast to an
analytical, spherical NFW profile.  Interestingly, however, the figure
shows that the average profile after stacking appears to remarkably well
match the NFW profile; the stacked profile and the best-fit NFW profile
are almost indistinguishable, on top of each other. For the sake of
comparison we plot the amplitudes and the radius relative to the
best-fit NFW model of the stacked profile as in
Figure~\ref{fig:test_nfw}: the best-fit parameters are $M^{\rm
bf}_{500c}\simeq 4\times 10^{14}~h^{-1}M_\odot$ and $c^{\rm
bf}_{500c}\simeq 2.50$. These numbers are compared to the averages of
their underlying true values: $\ave{M^{\rm 3D~fit}_{500c}}\simeq
4.38\times 10^{14}~h^{-1}M_\odot$, $\ave{c^{\rm 3D~fit}_{500c}}\simeq
2.76$ or $\ave{M^{\rm 2D~fit}_{500c}}\simeq 4.56\times
10^{14}h^{-1}M_\odot$ and $\ave{c^{\rm 2D~fit}_{500c}}\simeq 2.57$.  Thus the stacked
lensing tends to underestimate the true mass, confirming the claims in
the previous work
\citep{Mandelbaumetal:05,Meneghettietal:10,BeckerKravtsov:11,vandenBoschetal:13,Meneghettietal:14}.

The other three panels show the results with NFW scaling
implementation. The lensing profiles of individual halos or the stacked
lensing profile are estimated by summing the ``scaled'' amplitude of
lensing distortion in each of the ``scaled'' radial bin relative to the
NFW predictions of each halo (see Eq.~\ref{eq:est_fnfw}).  The different
panels are the results when using the best-fit NFW parameters of 3D mass
density profile for each halo, the NFW parameters of 2D lensing profile,
or the halo mass of 3D profile, but using the concentration parameter
inferred from the scaling relation, $c_{500c}=c(M_{500c})$ in DR14,
respectively\footnote{For the measurement we used the X-ray observables
to infer the halo mass of each cluster. Here we assumed that the X-ray
observables are sensitive to the inner region of each cluster or
relatively less sensitive to the projection effect than in weak
lensing. Hence we assume that the X-ray observables gives a proxy of the
halo mass via the 3D profile.}. The lower-right panel is closest to our
main results using the Subaru and X-ray data.  All the three panels
clearly show that the NFW scaling significantly reduces the scatters of
individual lensing profiles relative to the NFW prediction, compared to
the standard lensing analysis (the upper-left panel). The scatters
appear to be smallest when using the best-fit NFW parameters of the 2D
lensing profile. Comparing the upper-right and lower-right panels
clarifies how the scatters are enlarged due to the lack of halo
concentration knowledge on individual halo or shows the degradation by
ignoring the intrinsic scatters of halo concentration in different
halos. Encouragingly the figure shows that, even without knowledge on
the concentration of each halo, the NFW scaling can reduce the scatters
compared to the upper-left panel.  However it should be noted that the
stacked lensing profile shows a sizable deviation from the NFW profile
(the blue solid curve), compared to the upper-left panel. In summary
these results justify our approach of using the scatters of individual
lensing profiles in order to test the universality of mass density
profile in 50 clusters.

\section{Application}
\label{sec:results}

In this section we apply, as a proof of concept, the method developed in
the preceding section to the Subaru data for a sample of 50 massive
clusters.

\subsection{The cluster sample}
\label{sec:sample}

 \begin{table*}[H]
  \begin{center}
    \caption{X-ray and Lensing Observables of 50 Clusters \label{tab:x-ray}}
   \begin{tabular}{lccccccccc}
      \hline \hline
   \multicolumn{1}{l}{Cluster} & \multicolumn{1}{l}{redshift}  &
    \multicolumn{3}{c}{X-ray data} & \multicolumn{2}{c}{$M_{500c}$
    [$10^{14}M_{\odot}$]} & \multicolumn{3}{c}{Lensing observables}\\
	 & & $r_{500c}$ [Mpc] &Telescope & $M_{\mathrm{gas},500}~
		    [10^{14}M_\odot]$
		    &$M^\mathrm{HSE}_{500c}$ &
    $M^{M_\mathrm{gas}}_{500c}$ &
    $({S/N})_\mathrm{WL}$ & $(d^2)^\mathrm{HSE}$ & $(d^2)^{M_\mathrm{gas}}$\\
 \hline 
A$2697$ &$0.232$ &$1.20\pm0.04$ &{\it XMM} &$0.880\pm0.037$   &$6.29\pm0.65$ &$5.98\pm0.25$  & $6.42$ & $4.47$  & $4.62$  \\
 A$68$ &$0.255$ &$1.40\pm0.20$ &{\it XMM} &$0.903\pm0.135$   &$10.44\pm0.86$ &$6.13\pm0.92$ & $6.24$ & $13.62$ & $7.52$  \\
 A$2813$ &$0.292$ &$1.25\pm0.10$ &{\it XMM} &$1.010\pm0.092$   &$6.32\pm0.69$ &$6.86\pm0.63$  & $5.45$ & $4.66$  & $4.28$  \\
 A$115$ &$0.197$ &$0.89\pm0.07$ &{\it XMM} &$0.546\pm0.089$   &$3.26\pm0.21$ &$3.71\pm0.60$  & $3.61$ & $8.20$  & $8.46$  \\
 A$141$ &$0.230$ &$1.02\pm0.12$ &{\it Chandra}	  &$0.550\pm0.084$ &$3.97\pm1.37$ &$3.74\pm0.58$      & $5.35$ & $6.01$  & $6.36$  \\
 ZwCl$0104$ &$0.254$ &$0.76\pm0.01$ &{\it Chandra} &$0.235\pm0.004$ &$1.67\pm0.07$ &$1.60\pm0.03$      & $3.55$ & $5.46$  & $5.50$  \\
 A$209$ &$0.206$ &$1.15\pm0.07$ &{\it XMM} &$0.972\pm0.094$  &$5.45\pm0.18$ &$6.60\pm0.64$   & $9.14$ & $21.84$ & $17.37$ \\
 A$267$ &$0.230$ &$1.17\pm0.12$ &{\it Chandra}	  &$0.703\pm0.094$  &$5.97\pm1.84$ &$4.78\pm0.64$     & $5.37$ & $7.01$  & $7.47$  \\
 A$291$ &$0.196$ &$0.94\pm0.06$ &{\it XMM} &$0.391\pm0.031$  &$2.92\pm0.56$ &$2.66\pm0.21$   & $4.92$ & $8.61$  & $8.92$  \\
 A$383$ &$0.188$ &$1.01\pm0.08$ &{\it XMM} &$0.425\pm0.036$   &$3.25\pm0.45$ &$2.89\pm0.24$  & $5.74$ & $9.11$  & $9.80$  \\
 A$521$ &$0.248$ &$1.25\pm0.04$ &{\it XMM} &$1.230\pm0.050$  &$7.05\pm0.34$ &$8.36\pm0.34$   & $5.98$ & $4.66$  & $5.23$  \\
 A$586$ &$0.171$ &$1.09\pm0.08$ &{\it Chandra}	  &$0.600\pm0.049$  &$4.42\pm0.90$ &$4.08\pm0.33$     & $5.58$ & $10.90$ & $11.47$ \\
 A$611$ &$0.288$ &$1.20\pm0.06$ &{\it Chandra}	  &$0.612\pm0.039$  &$6.80\pm1.08$ &$4.16\pm0.26$     & $5.98$ & $8.24$  & $12.30$ \\
 A$697$ &$0.282$ &$1.50\pm0.10$ &{\it Chandra}	  &$1.490\pm0.127$  &$13.14\pm2.74$ &$10.12\pm0.86$    & $4.89$ & $20.89$ & $12.84$ \\
 ZwCl$0857$ &$0.235$ &$0.85\pm0.04$ &{\it XMM} &$0.351\pm0.017$ &$2.33\pm0.23$ &$2.38\pm0.12$ & $2.45$ & $4.53$  & $4.60$  \\
 A$750$ &$0.163$ &$0.97\pm0.12$ &{\it Chandra}	  &$0.406\pm0.061$  &$3.17\pm1.22$ &$2.76\pm0.41$     & $6.82$ & $16.54$ & $18.17$ \\
 A$773$ &$0.217$ &$1.21\pm0.11$ &{\it XMM} &$0.907\pm0.102$  &$7.77\pm0.94$ &$6.16\pm0.69$   & $9.03$ & $7.40$  & $11.40$  \\
 A$781$ &$0.298$ &$1.13\pm0.17$ &{\it XMM} &$0.783\pm0.142$  &$6.72\pm0.45$ &$5.32\pm0.96$   & $4.27$ & $28.93$ & $25.32$ \\
 ZwCl$0949$ &$0.214$ &$0.93\pm0.02$ &{\it Chandra} &$0.315\pm0.010$  &$2.90\pm0.20$ &$2.14\pm0.07$     & $5.50$ & $15.34$ & $16.79$ \\
 A$901$ &$0.163$ &$0.79\pm0.06$ &{\it XMM} &$0.208\pm0.020$ &$1.68\pm0.39$ &$1.41\pm0.14$    & $5.58$ & $11.98$ & $13.43$ \\
 A$907$ &$0.167$ &$1.08\pm0.06$ &{\it XMM} &$0.623\pm0.042$ &$5.17\pm0.64$ &$4.23\pm0.29$    & $7.71$ & $11.32$ & $14.53$ \\
 A$963$ &$0.205$ &$1.14\pm0.07$ &{\it XMM} &$0.616\pm0.050$  &$5.60\pm0.71$ &$4.18\pm0.34$   & $7.49$ & $13.40$ & $17.46$ \\
 ZwCl$1021$ &$0.291$ &$1.26\pm0.05$ &{\it XMM} &$1.080\pm0.048$ &$6.82\pm0.14$ &$7.34\pm0.33$ & $6.78$ & $8.76$  & $9.23$ \\ 
 A$1423$ &$0.213$ &$1.18\pm0.10$ &{\it Chandra}	  &$0.711\pm0.095$ &$6.02\pm1.53$ &$4.83\pm0.65$      & $4.58$ & $9.89$  & $6.95$  \\
 A$1451$ &$0.199$ &$1.36\pm0.11$ &{\it XMM} &$1.050\pm0.102$  &$8.97\pm2.18$ &$7.13\pm0.69$   & $8.25$ & $6.30$  & $5.44$  \\
 RXCJ$1212$ &$0.269$ &$0.76\pm0.05$ &{\it XMM} &$0.196\pm0.012$ &$1.67\pm0.31$ &$1.33\pm0.08$ & $3.00$ & $9.34$  & $8.86$  \\
 ZwCl$1231$ &$0.229$ &$1.23\pm0.08$ &{\it Chandra} &$0.828\pm0.078$  &$6.82\pm1.29$ &$5.63\pm0.53$     & $5.09$ & $17.35$ & $16.05$ \\
 A$1682$ &$0.226$ &$1.24\pm0.18$ &{\it Chandra}	  &$0.764\pm0.137$  &$7.35\pm3.06$ &$5.19\pm0.93$     & $7.30$ & $6.12$  & $8.03$  \\
 A$1689$ &$0.183$ &$1.52\pm0.07$ &{\it XMM} &$1.290\pm0.059$  &$11.98\pm1.94$&$8.76\pm0.40$   & $9.42$ & $11.25$ & $16.82$ \\
 A$1758$N &$0.280$ &$1.38\pm0.07$&{\it Chandra}	  &$1.220\pm0.062$ &$10.21\pm1.54$&$8.29\pm0.42$      & $3.10$ & $13.65$ & $10.03$ \\
 A$1763$ &$0.228$ &$1.33\pm0.11$ &{\it XMM} &$1.230\pm0.135$  &$6.60\pm0.56$ &$8.36\pm0.92$   & $6.75$ & $9.58$  & $7.09$  \\
 A$1835$ &$0.253$ &$1.57\pm0.11$ &{\it XMM} &$1.550\pm0.120$ &$14.04\pm1.27$ &$10.53\pm0.82$   & $5.96$ & $14.77$ & $13.27$ \\
 A$1914$ &$0.171$ &$1.38\pm0.08$ &{\it XMM} &$1.160\pm0.073$  &$8.08\pm1.00$ &$7.88\pm0.50$   & $4.83$ & $9.22$  & $8.88$ \\
 ZwCl$1454$ &$0.258$ &$1.06\pm0.10$ &{\it XMM} &$0.578\pm0.060$ &$3.65\pm0.42$ &$3.93\pm0.41$ & $3.18$ & $3.94$  & $4.13$  \\
 A$2009$ &$0.153$ &$1.29\pm0.13$ &{\it Chandra}	  &$0.708\pm0.082$  &$7.33\pm2.47$ &$4.81\pm0.56$     & $4.91$ & $5.79$  & $5.14$  \\
 ZwCl$1459$ &$0.290$ &$1.08\pm0.25$ &{\it XMM} &$0.675\pm0.187$ &$5.65\pm0.36$ &$4.59\pm1.27$ & $3.31$ & $4.16$  & $3.73$  \\
 RXCJ$1504$ &$0.215$ &$1.47\pm0.35$ &{\it XMM} &$1.300\pm0.337$ &$10.93\pm0.82$ &$8.83\pm2.29$& $4.55$ & $7.70$  & $6.23$  \\
 A$2111$ &$0.229$ &$1.17\pm0.14$ &{\it Chandra}	  &$0.719\pm0.110$  &$5.99\pm1.89$ &$4.88\pm0.75$     & $4.83$ & $13.10$ & $12.57$ \\
 A$2204$ &$0.152$ &$1.49\pm0.08$ &{\it XMM} &$1.280\pm0.082$ &$10.66\pm1.72$ &$8.70\pm0.56$   & $6.48$ & $11.20$ & $9.16$ \\
 A$2219$ &$0.228$ &$1.75\pm0.11$ &{\it XMM} &$1.882\pm0.216$
    &$14.35\pm2.04$ &$12.79\pm1.47$   & $7.51$ & $6.21$  & $6.06$  \\
 RXCJ$1720$ &$0.164$ &$1.23\pm0.11$ &{\it XMM} &$0.771\pm0.083$ &$6.97\pm0.68$ &$5.24\pm0.56$ & $3.80$ & $11.47$ & $7.02$  \\
 A$2261$ &$0.224$ &$1.22\pm0.12$ &{\it Chandra}	  &$1.000\pm0.127$  &$6.75\pm1.89$ &$6.79\pm0.86$     & $8.88$ & $13.44$ & $13.33$ \\
 RXCJ$2102$ &$0.188$ &$1.00\pm0.06$ &{\it XMM} &$0.450\pm0.033$ &$3.52\pm0.61$ &$3.06\pm0.22$ & $4.04$ & $13.87$ & $12.78$ \\
 RXJ$2129$ &$0.235$ &$1.08\pm0.04$ &{\it XMM} &$0.749\pm0.037$  &$4.22\pm0.16$ &$5.09\pm0.25$ & $3.17$ & $5.10$  & $6.17$  \\
 A$2390$ &$0.233$ &$1.60\pm0.11$ &{\it XMM} &$1.700\pm0.088$  &$13.67\pm2.09$ &$11.55\pm0.60$  & $6.30$ & $8.47$  & $7.41$  \\
 A$2485$ &$0.247$ &$1.11\pm0.15$ &{\it Chandra}	  &$0.558\pm0.087$  &$5.32\pm2.08$ &$3.79\pm0.59$     & $4.71$ & $1.89$  & $0.52$  \\
 A$2537$ &$0.297$ &$1.19\pm0.10$ &{\it XMM} &$0.739\pm0.081$   &$7.20\pm0.73$ &$5.02\pm0.55$  & $5.05$ & $8.73$  & $9.81$  \\
 A$2552$ &$0.300$ &$1.25\pm0.09$ &{\it Chandra}	  &$1.020\pm0.094$  &$7.81\pm1.64$ &$6.93\pm0.64$     & $3.98$ & $11.30$ & $10.85$ \\
 A$2631$ &$0.278$ &$1.20\pm0.09$ &{\it XMM} &$1.030\pm0.088$  &$8.51\pm0.98$ &$6.93\pm0.64$   & $4.61$ & $18.05$ & $16.23$ \\
 A$2645$ &$0.251$ &$1.15\pm0.18$ &{\it Chandra}	  &$0.541\pm0.117$  &$5.98\pm2.59$ &$3.68\pm0.79$     & $6.59$ & $19.54$ & $19.62$ \\
  \hline\hline
   \end{tabular}
  \end{center}
  \tablecomments{The X-ray observables ($r_{500c}$, $M_{500c}^{\rm HSE}$
  and $M_{\rm gas, 500}$) taken from Tables~2 and 3 of
  \citet{Martinoetal:14}:  
  $r_{500c}$ is the radius for the interior overdensity
  $\Delta=500$, $M_{500c}^{\rm HSE}$ is the mass estimate based on the
  hydrostatic equilibrium, and $M_{\rm gas, 500}$ is the gas mass
  interior to $r_{500}$ (see text for details).
  $M^{M_{\rm gas}}_{500c}$ is the total mass interior to $r_{500c}$
  assuming the {\it simple} self-similar scaling relation given by
  Eq.~(\ref{eq:M500-Mgas}). The mean mass of 50 clusters
  $\ave{M_{500c}}/[10^{14}h^{-1}M_\odot]=4.42$ or $3.82$ for the HSE and
  gas mass cases, respectively.
  The last three columns are the lensing
  observables that are computed from the lensing measurement of
  \citet{Okabeetal:13}. $(S/N)_{\rm WL}$ is the total signal-to-noise
  ratio of lensing distortion measurement for each cluster over the 8
  radial bins in the range $0.14\le R/[h^{-1}{\rm Mpc}]\le
  2.8$. $(d^2)^{\rm HSE}$ or $(d^2)^{M_{\rm gas}}$ is the deviation of
  the lensing distortion profile compared to the NFW prediction, defined
  by Eq.~(\ref{eq:d}) or (\ref{eq:d_nfw}) for each cluster.}
 \end{table*}

\subsubsection{Subaru weak lensing data}
\label{sec:subaru}

For the weak lensing measurements, we use the shape catalog of galaxies
for the 50 clusters, used in the published work of \citet{Okabeetal:13}.
This is the older version of shape catalog, derived as a part of the
LoCuSS collaboration \citep[see][for
details]{Okabeetal:10,Martinoetal:14}.  In brief, the 50 cluster sample
was selected from the {\it ROSAT} All Sky Survey catalogs
\citep{Ebelingetal:98, Ebelingetal:2000, Bohringeretal:04} that satisfy
the criteria given as $L_X[0.1-2.4{\rm keV}]/E(z)^{2.7}\ge
4.2\times10^{44}\,{\rm erg\,s^{-1}}$, $0.15\le z\le 0.30$,
$n_H<7\times10^{20}{\rm cm^{-2}}$, and $-25^\circ<\delta<+65^\circ$,
where $E(z)\equiv H(z)/H_0$ is the normalized Hubble expansion rate.
The criteria on the redshift range and the declination are adopted in
order to have a sufficiently high elevation of these clusters from the
Subaru telescope and to have an entire coverage of the virial region of
these clusters with the field of view of the Subaru Suprime-Cam camera
\citep{Miyazakietal:02}.

All the clusters were observed by Subaru, with two passbands at least:
$i$ or $I_C$ data, which was used for the weak lensing analysis in
\citet{Okabeetal:13},
and the
bluer-passband data, $V$ or $g$ data.
For this paper, we
take the position of brightest cluster galaxy in each cluster as the
cluster center. \citet{Okabeetal:10} carefully studied a possible
miscentering effect by comparing the lensing signals of various center
proxies such as the X-ray peak, and concluded that the miscentering,
even if exists, should be well within 100~kpc in radius (more exactly,
within about 50~kpc in our estimate), which is inside the minimum radius
used in this paper.

An important systematic effect in the weak lensing measurements is a
possible residual uncertainty in estimation of source galaxy redshifts,
mainly limited by the two passband data alone. \citet{Okabeetal:13}
developed a method of making a secure sample of background galaxies,
which is selecting galaxies with color sufficiently redder than the
red-sequence of early-type galaxies in each cluster region. In other
words, they found that it is very difficult to select ``blue''
background galaxies from the two passband data alone or such blue
galaxies always appear to be contaminated by foreground or member
(therefore unlensed) galaxies.  However, this selection is conservative
and leaves only a small number of galaxies in the sample so as to ensure
less than 1$\%$ contamination or dilution effect on the lensing signal,
even if exists: the mean number density of galaxies is about 5
arcmin$^{-2}$, a factor 4 or 5 smaller than the number density of all
the galaxies for which  weak lensing analysis is usable in the original $i$- or
$I_c$-band catalog. Hence the measurement errors of weak lensing signals
are dominated by the shape noise, which justifies that we ignore the
error contribution of projection effects due to different structures
along the same line-of-sight to the cluster. The mean redshift of
background galaxies in each cluster was estimated by matching color of
the selected background galaxies to the COSMOS catalog. Since all the
clusters are at low redshift $z\sim 0.2$ and the deep Subaru data
typically probe galaxies at $z\sim 0.8$, the lensing efficiency has a
weak dependence on source redshift and a possible residual uncertainty
in the source redshift would not be large and should be less than a 10\%
change in the lensing amplitude even if exists \citep[see \S~5.7.2
in][]{Okabeetal:10}. We should also keep in mind an additional
uncertainty due to the sample variance in the COSMOS calibration
catalog, which refers a possible difference in the populations of source
galaxies in between the COSMOS and cluster regions.

Again note that the purpose of this paper is to give a proof of concept
of the novel cluster lensing measurement method, so the results we will
show below is based on the catalog of \citet{Okabeetal:13}. See Okabe et
al. (2015) for the improved results of weak lensing measurements based
on a more careful treatment of shape measurement and photo-$z$
uncertainty.

\subsubsection{X-ray observables: hydrostatic equilibrium mass, gas mass
  and gas temperature}
\label{sec:x-ray}

 \begin{figure*}
  \centering{
  \includegraphics[width=0.7\textwidth]{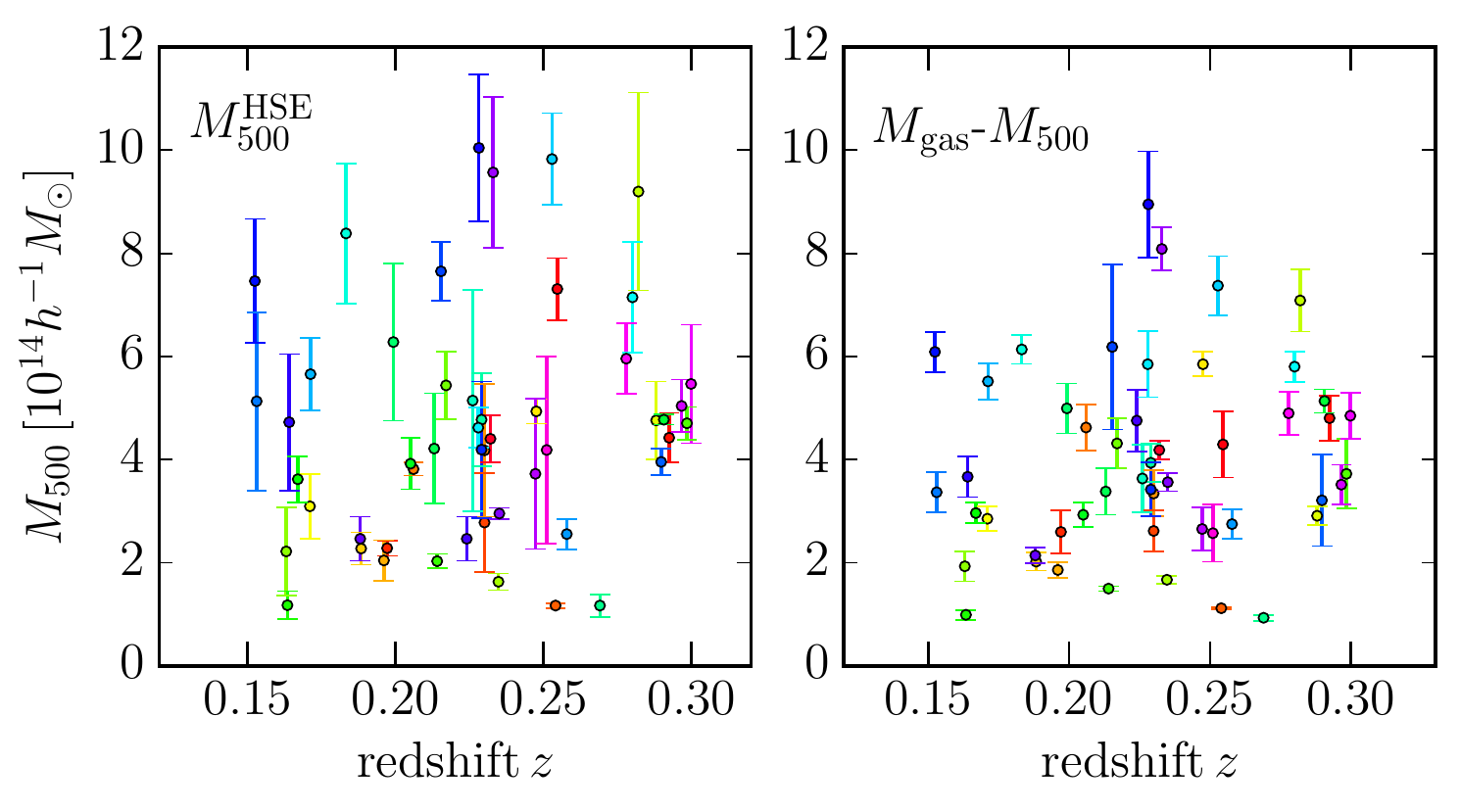}}
\caption{The 50 clusters used in this paper, in the cluster mass
  ($M_{500c}$ ) and redshift plane.  The mass estimate of each cluster
  was taken from \citet{Martinoetal:14} (see also
  Table~\ref{tab:x-ray}), derived based on the {\it Chandra} and/or {\it
  XMM} X-ray data.  The left panel is the mass estimate based on the
  hydrostatic equilibrium (HSE) assumption, while the right panel shows
  the results derived using the scaling relation of X-ray gas mass with
  halo mass (Eq.~\ref{eq:M500-Mgas}), respectively. The errorbars of
  each cluster are also taken from \cite{Martinoetal:14} (for
  $M_{\rm gas, 500}$ we propagated the errors of gas mass). Comparing
  the two panels reveals that the same cluster (symbols at the same
  $x$-axis value) generally has different mass estimates and errorbars.
  } \label{fig:m-z}
\end{figure*}
All the 50 clusters were observed by the X-ray satellites, {\it
XMM}-{\it Newton} or/and {\it Chandra}
\citep{Zhangetal:10,Martinoetal:14}.  In this paper, we use the X-ray
observables in \citet{Martinoetal:14} to infer the halo mass for each of
the 50 clusters, which was estimated based on either or both of the {\it
XMM} and/or {\it Chandra} data. In the following we will use two mass
estimates: the mass estimated based on the hydrostatic equilibrium
assumption (hereafter HSE for simplicity) and the self-similar scaling
relation of gas mass ($M_{\rm gas}$) with the halo mass:
 \begin{itemize}
  \item {\it Hydrostatic equilibrium (HSE) mass} --
	\citet{Martinoetal:14} carefully developed a method of
	estimating the HSE mass of each cluster by combining the surface
	brightness and temperature profiles, measured from the {\it Chandra}
	and/or {\it XMM} data:
\begin{equation}
 M(<r)=-\frac{kT_{g}(r)r}{G\mu m_p}\left[
\frac{d \ln\rho_g(r)}{d \ln r}+\frac{d\ln T_g(r)}{d\ln r}
							     \right],
 \label{eq:hse}
\end{equation}
	where $T_g(r)$ and $\rho_g(r)$ are the three-dimensional radial
	profiles of gas temperature and density, respectively. We will
	use the halo mass estimate for $\Delta=500$ overdensity,
	$M_{500c}$, in Table~2 of \citet{Martinoetal:14}. For some of
	the clusters, the mass estimates were derived for both the
	{\it Chandra} and {\it XMM} data. If the two mass estimates are available,
	we use the {\it XMM}-derived mass because the mass accuracy is better
	than that of the {\it Chandra}-based estimate. Note that the two
	estimates are consistent with each other within the errorbars,
	We use the {\it XMM}-based mass for 32 clusters, and use the
	{\it Chandra}-based mass for the remaining 18 clusters.
  \item {\it $M_{\rm gas}$ derived mass} -- The direct X-ray observables are
	the gas mass and temperature. If non-gravitational processes are
	not significant for cluster evolution, the mass, temperature,
	size and other properties of galaxy cluster follow self-similar
	scaling relations \citep{Kaiser:86}.  The ratio of the total
	matter and gas masses in a cluster region is expected to follow
	the scaling relation:
	%
	 $M_\Delta(<r) \propto M_{{\rm gas},\Delta}(<r).$
	%
	For the interior gas mass, we will use the $M_{{\rm gas},500}$
	value in Table~3 of \citet{Martinoetal:14} for each cluster. For
	the normalization factor, we here simply employ the cosmic mean
	value that is inferred from the latest Planck result \citep{PlanckCosmo:15}: 
	\begin{eqnarray}
	 \frac{M_{\rm
	  500c}}{10^{14}M_\odot}&=&\frac{\Omega_{\rm m0}}{\Omega_{\rm
	  b0}}\frac{M_{\rm gas, 500}}{10^{14}M_\odot}\nonumber\\
	  &\simeq & 11.6\times
	 \left(\frac{M_{\rm
	  gas,500}}{10^{14}h^{-3/2}M_\odot}\right),
	 \label{eq:M500-Mgas}
	\end{eqnarray}
	where we took the best-fit values of $\Omega_{\rm b0}h^2$,
	$\Omega_{\rm m0}h^2$ and $h$ in Table 3 of
	\citet{PlanckCosmo:15} to compute the normalization constant.
	The unit $h^{-3/2}$ of gas mass is from the fact that the gas
	mass estimate from X-ray observables has the $h$-dependence.
	Note that the overdensity radius $r_{500c}$ used for the
	interior mass definition is from the total mass profile derived
	from the HSE assumption, Eq.~(\ref{eq:hse}).  In this sense,
	exactly speaking, this treatment is not self-consistent.
	Comparing the above normalization constant with Figure~2 in
	\citet{Okabeetal:14} shows that our model is within a range of
	the normalization constants implied from observations. However,
	a precise determination of the normalization constant is not our
	primary purpose, and the above choice can be considered as a
	working example. We will below study how variations in the above
	scaling relation change the weak lensing measurements with NFW
	scaling.
\end{itemize}
Table~\ref{tab:x-ray} gives a summary of the above X-ray observables:
the HSE mass and the gas mass for each. Figure~\ref{fig:m-z} shows the
distribution of 50 clusters in the plane of halo mass and redshift. The
two proxies give a different estimate of halo mass on individual cluster
basis and the errorbars quoted are also different.
The mean mass of 50 clusters (without lensing weights),
$\ave{M_{500c}}/[10^{14}h^{-1}M_\odot]=4.42$ or 3.82 for the HSE or gas
mass proxy, respectively.

\subsection{The stacked lensing analysis of 50 clusters with and
 without NFW scaling}
\label{sec:fnfw}

\begin{figure}
\centering{
\includegraphics[scale=0.45]{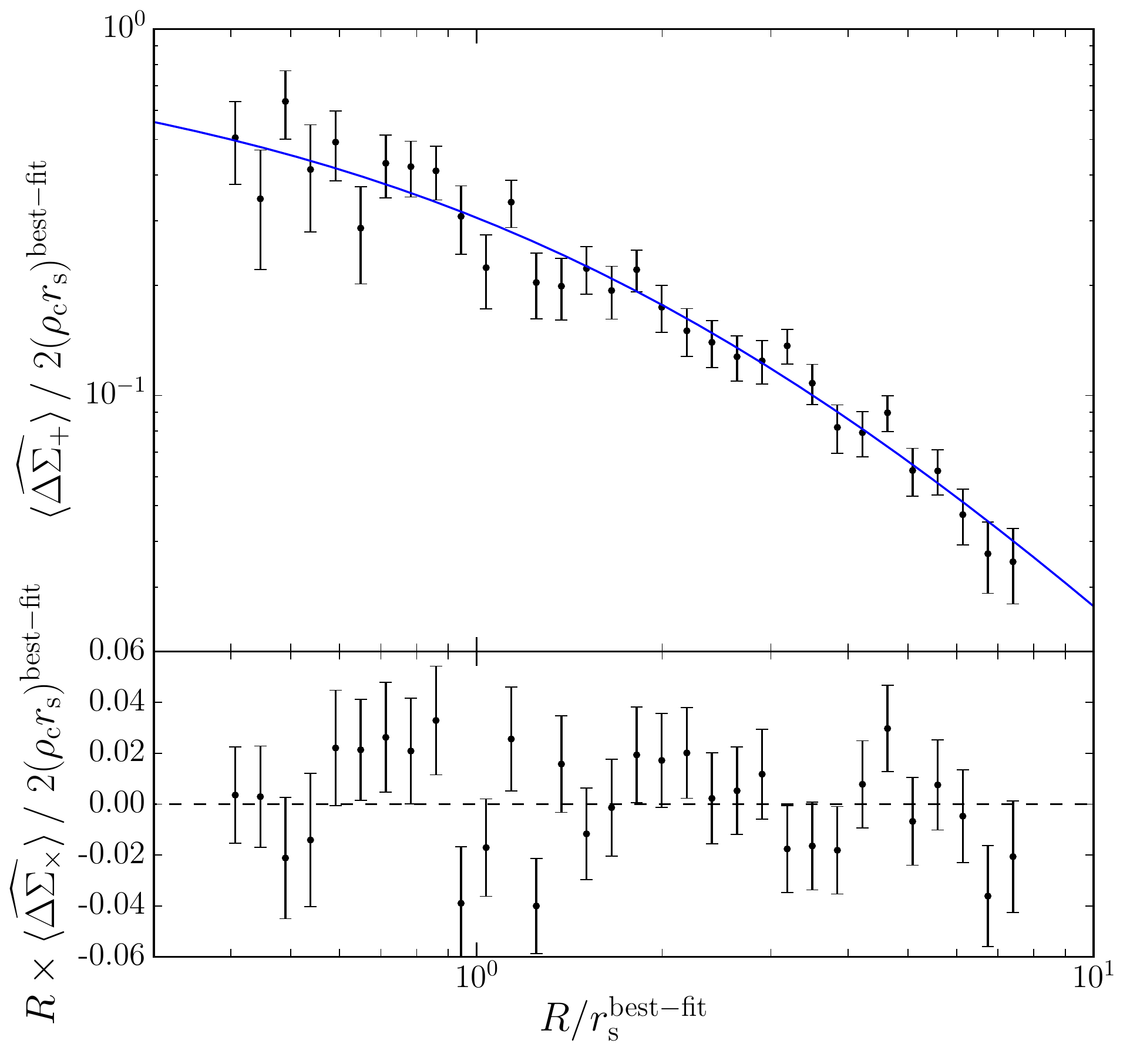}}
 \caption{{\it Upper panel}: The stacked distortion profile measured
 from $50$ Subaru clusters, based on the standard stacked lensing
 analysis (Eq.~\ref{eq:est_dSigma}), i.e. without NFW scaling
 implementation. We employed the 32 logarithmically-spaced bins over a
 fixed range of radii, $0.14\le R/[h^{-1}\mbox{Mpc}]<2.8$, for all the
 50 clusters.
 The errorbar at each bin is computed from Eq.~(\ref{eq:e_est_dSigma})
 assuming that the statistical noise is dominated by the intrinsic
 ellipticities of background galaxies.  The solid curve is the best-fit
 NFW model, which is specified by the best-fit parameters
 $M_{500c}=(4.0\pm0.1)\times 10^{14}h^{-1}{M_\odot}$ and $c_{500c}= 2.8
 \pm 0.3$. The best-fit model reproduces the results in
 \citet{Okabeetal:13} within the errorbars\footnote{\citet{Okabeetal:13}
 used the physical coordinates in the stacking analysis, and this leads
 to a slight change in the best-fit NFW parameters.}. 
 For the sake of comparison with the following figures, we
 plot the distortion profile relative to the best-fit NFW model, as a
 function of the radius relative to the scale radius of the best-fit NFW
 model.  Note that the representative value of each radial bin is
 estimated from the average of radii of background galaxies that reside
 in the annulus (see Eq.~\ref{eq:r}). The reduced chi-square for the
 best-fit model is $\chi^2/{\rm d.o.f}=22.5/(32-2)$.  {\it Lower panel}:
 Similar to the above panel, but for the $45^{\circ}$-rotated components
 of background galaxy ellipticities.  } \label{fig:dSigma}
\end{figure}
First of all, in Figure~\ref{fig:dSigma}, we show the stacked lensing
profile of 50 clusters, without NFW scaling, for the sake of comparison
with the following results. This result reproduces Figure~3 in
\citet{Okabeetal:13}.  We employed 32 logarithmically-spaced bins over
the radial range of $0.14\le R/[h^{-1}{\rm Mpc}] \le 2.8$. As given by
Eq.~(\ref{eq:r}), we estimated the representative value of each radial
bin by averaging the centric-radii of background galaxies in the
annulus, and therefore the neighboring bins are, exactly speaking, not
equally spaced, although the difference is very small after the average
of 50 clusters.  The cumulative signal-to-noise ratio is significant:
$S/N\simeq 34.5$. From the fitting to an NFW profile, we find the
best-fit parameters, $M_{500c}=(4.0\pm0.1)\times 10^{14}h^{-1}{M_\odot}$
and $c_{500c}= 2.8 \pm 0.3$, respectively. The reduced chi-square is
$\chi^2/{\rm d.o.f}=22.5/(32-2)$.  Thus the results show that, even if
the X-ray inferred masses differ from each other by up to a factor of
10, the stacked profile is so remarkably well fitted by the NFW
model. This appears to be consistent with what we found from the test
using the simulated halos in Figure~\ref{fig:test_nbody}.

\begin{figure}
 \centering{
 \includegraphics[width=0.45\textwidth]{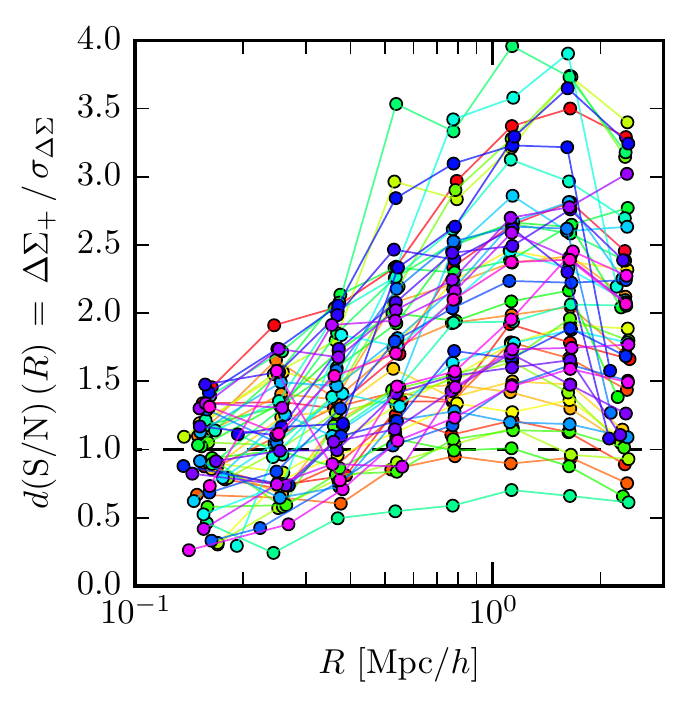}
 \caption{The expected, differential signal-to-noise ratio, ${\rm
 d}(S/N)$, at each of the 8 logarithmically-spaced radial bins in the
 range $0.14\le R/[h^{-1}{\rm Mpc}]\le 2.8 $, for each of 50 Subaru
 clusters. We computed the $d(S/N)$ value as follows. For the expected
 signal, we used an analytical NFW profile for each cluster assuming the
 X-ray HSE mass and the halo concentration inferred from the $c-M$
 relation in DK15. To compute the statistical noise in each bin, we used
 the real Subaru data of background galaxies (their distribution on the
 sky, the intrinsic shapes and the lensing weights) in each cluster
 region.
 Most of the data points, about 79\% of 400 data points ($400=50\times
 8$), are expected to have the ${\rm d}(S/N)$ values greater than unity. The same
 color symbols at different radial bins correspond to the same
 cluster. Note that the representative value of each radial bin is
 computed from Eq.~(\ref{eq:r}) taking into account the radii and
 weights of background galaxies, which causes variations in the
 representative values especially for the small radii, even if we work
 on the fixed range of $0.14\le R/[h^{-1}{\rm Mpc}]\le 2.8$.
 \label{fig:sn_8bin}}}
\end{figure}
\begin{figure*}
 \centering{
 \includegraphics[width=1.0\textwidth]{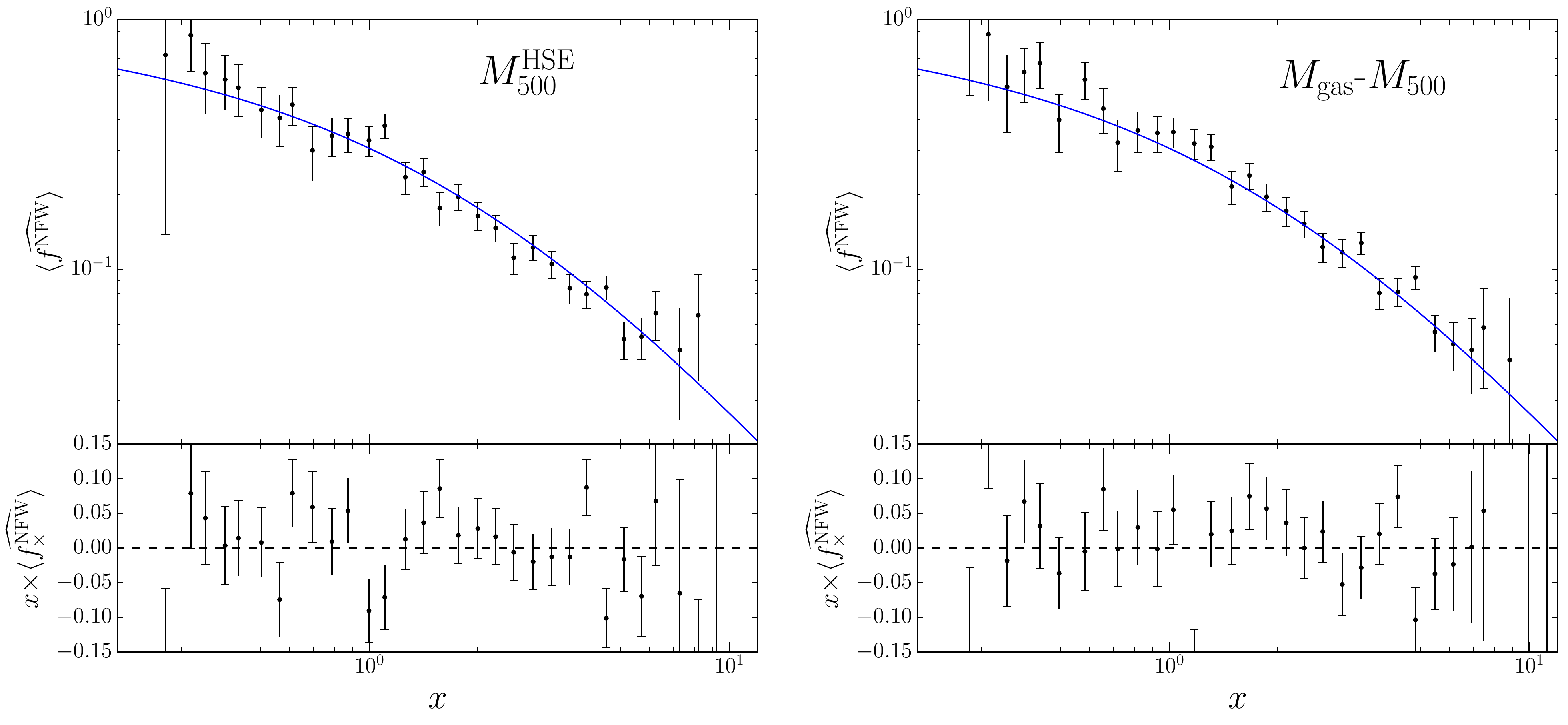} }
 \caption{Similar to Figure~\ref{fig:dSigma}, but the stacked distortion
 profile of 50 clusters when implementing the NFW scaling analysis: we
 summed the ``scaled'' amplitudes of background galaxy ellipticities in
 each bin of the ``scaled'' radii according to the NFW parameters, halo
 mass and concentration, inferred for each cluster based on its X-ray
 observables. In the left or right panels, we employed the X-ray
 inferred mass of each cluster from the hydrostatic equilibrium
 assumption (HSE) or the gas mass, respectively, and then used the halo
 concentration inferred from the scaling relation $c=c(M; z)$ in
 DK15. Note that we used exactly the same background galaxies as those
 for the analysis without NFW scaling in Figure~\ref{fig:dSigma}.
 The errorbar at each bin is computed based on
 Eq.~(\ref{eq:e_est_fnfw}). The solid curve in each panel is {\it not} a
 fit, but the NFW prediction ($f^{\rm NFW}$ given by Eq.~\ref{eq:fnfw})
 including a small correction due to reduced shear at the small radii
 (see below Eq.~\ref{eq:est_fnfw} for details). The reduced chi-square
 is $\chi^2/{\rm d.o.f}=31.3/32$ or $30.7/32$ for the HSE or gas mass
 case, respectively.
 \label{fig:fnfw}}
\end{figure*}
\begin{figure*}
\centering{
\includegraphics[width=0.9\textwidth]{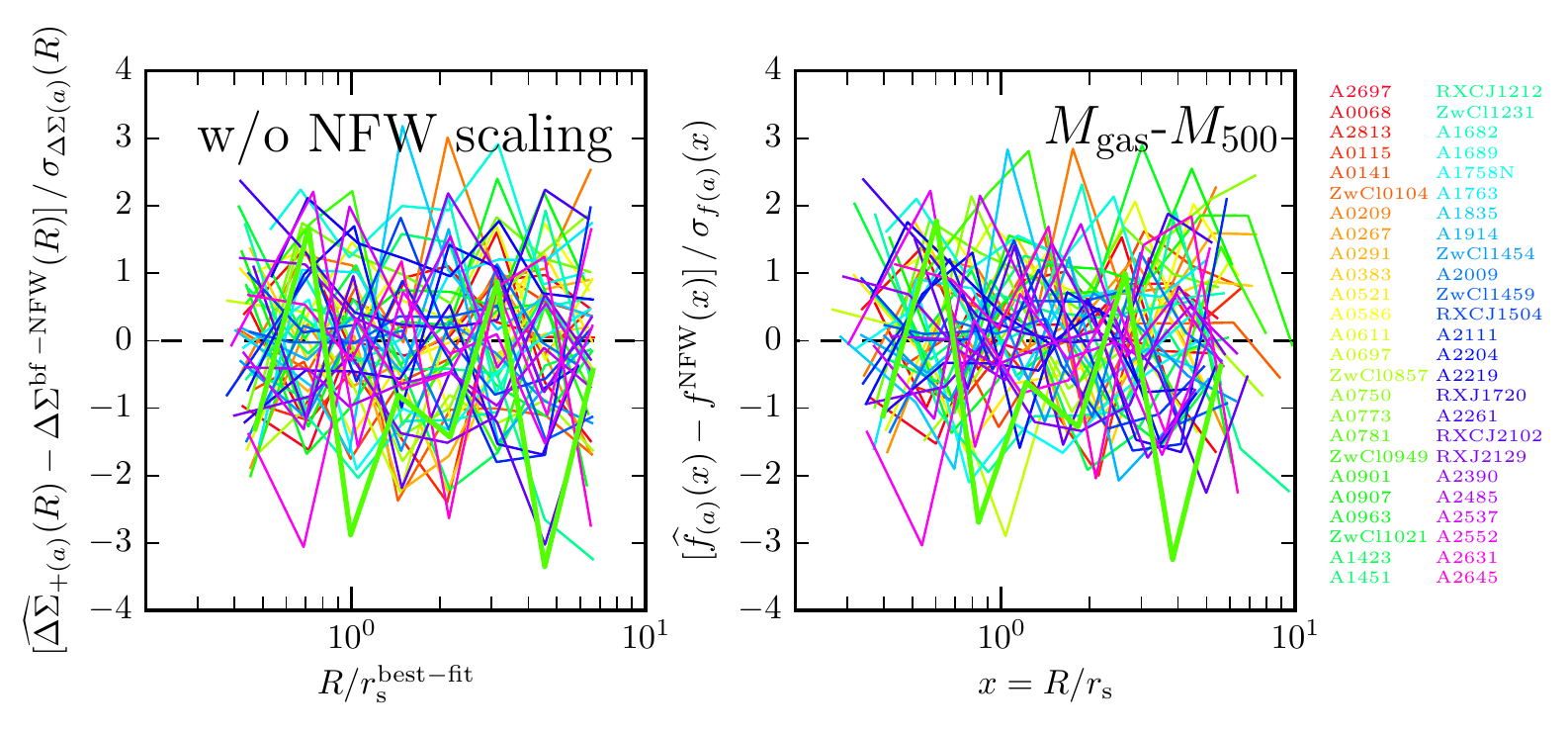} }
 \caption{The difference between the lensing distortion profiles of 50
 clusters and the best-fit NFW profile ($\Delta\Sigma^{\rm
 best-fit}(R)$) or the normalized NFW profile ($f^{\rm NFW}(x)$) for the
 weak lensing analysis with or without NFW scaling implementation in the
 left or right panel, respectively, as in the lower panels of
 Figure~\ref{fig:test_nbody}. Here we show in the right panel the result
 when using the gas mass to estimate the halo mass of each cluster, as
 in the right panel of Figure~\ref{fig:fnfw}. To make a fair comparison,
 we show the relative difference to the statistical error at each radial
 bin (see Eqs.~\ref{eq:d} and \ref{eq:d_nfw}).  Since the lensing
 profile is noisy on individual cluster basis due to the fewer number of
 background galaxies, we employed the 8 logarithmically-spaced bins in
 the fixed range of $0.14\le R/[h^{-1}{\rm Mpc}]\le 2.8$ for all the
 clusters as in Figure~\ref{fig:sn_8bin}. In addition, we used the same
 background galaxies in each radial bin before and after the NFW scaling
 transformation, $x=R/r_{s}$, for each cluster so that the differences
 become identical if we set the model NFW profile $\Delta\Sigma^{\rm
 bf-NFW}=f^{\rm NFW}=0$ (see the procedure 2b in \S~\ref{sec:fnfw} for
 details). Also note that, due to the NFW scaling, the fixed radial
 range in the left panel is transformed to the different range of the
 scaled radius for different clusters. The same-color curves in the two
 panels correspond to the same cluster, and the bold curve shows, as an
 example, the result for A781, which has the largest deviation from the
 NFW profile. Since the sum of squares of all the curves gives an
 estimate to quantify the scatters of 50 cluster lensing profiles
 relative to the NFW model -- we call the $d^2$ value. The NFW scaling
 yields $d^2=527.1$ or $504.6$ for the HSE and gas mass cases,
 respectively, compared to $d^2=543.2$ for the case without NFW scaling
 (Figure~\ref{fig:dSigma}). This corresponds to the improvement $\Delta
 d^2=d^2-d^2_{\rm w-scaling}=(4.0)^2 $ or $(6.2)^2$, respectively.
 \label{fig:scatters_data}}
\end{figure*}
\begin{figure}
 \centering{
 \includegraphics[width=0.49\textwidth]{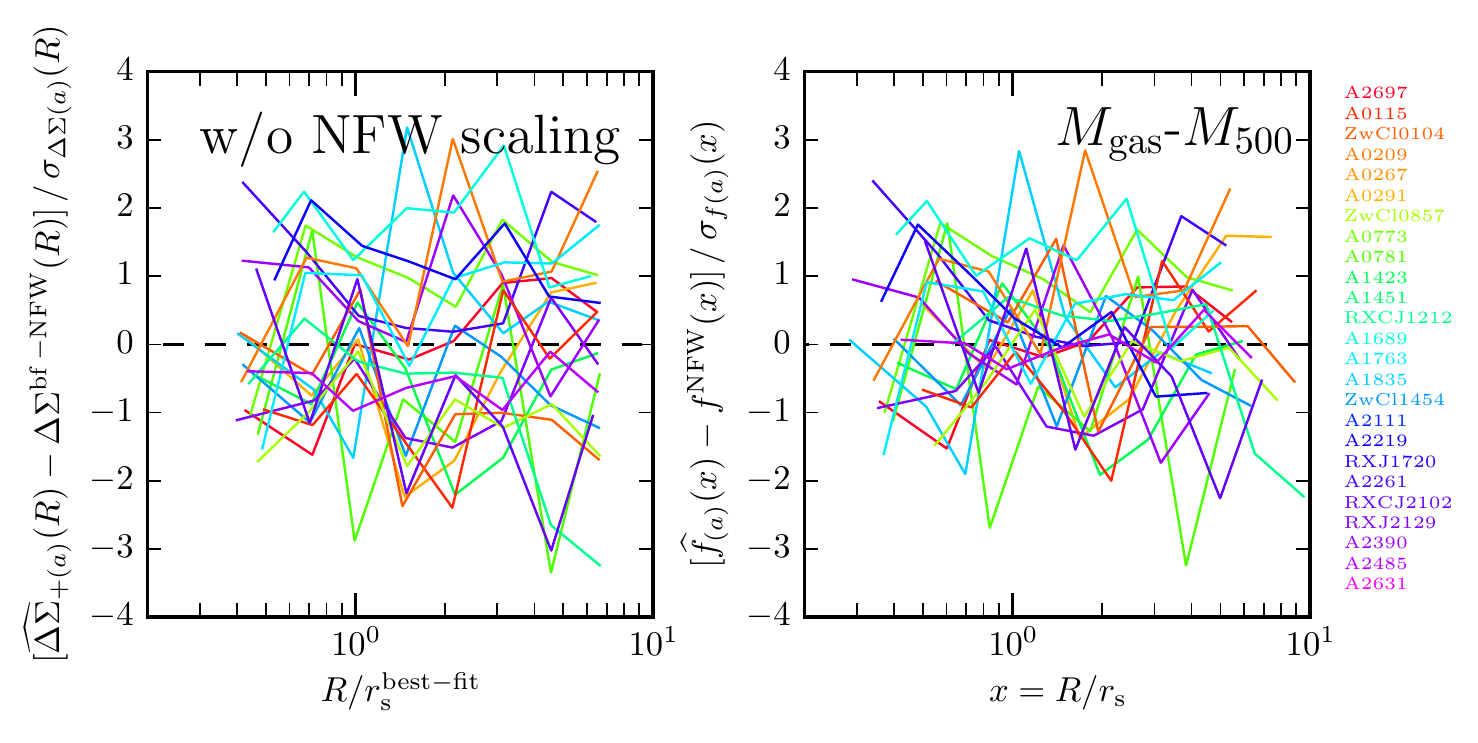}
 \caption{Similar to the previous figure, but the scatters for
 25 clusters, which are a half of 50 clusters that have the larger value of
 $\Delta d^2$ (Eq.~\ref{eq:dd2})
 than other 25 clusters 
 on individual cluster basis. Comparing the left and right panels more
 clearly shows that the NFW scaling reduces scatters of the lensing
 profiles. The cumulative value of $\Delta d^2$ for 25 clusters, 
 $\Delta d^2\simeq (9.3)^2$.
 \label{fig:scatters_cl25}}
 }
\end{figure}
We now move to the main results of this paper. We employ the
following procedures to implement the NFW scaling analysis of weak
lensing measurements: 
\begin{itemize}
 \item[(1)] {\it NFW scaling of galaxy ellipticities and radial bins} --
	    First, we employ, for the $a$-th cluster
	    ($a=1,2,\cdots,50$), the halo mass inferred from the X-ray
	    observables,
       either HSE or gas mass (see \S~\ref{sec:x-ray}). We then use the
	    $c$-$M$
       relation in DK15 to infer the halo
       concentration for the cluster. Using the X-ray inferred
       parameters, $M_{500(a)}^{\rm X}$ and
       $c^{\rm X}_{500(a)}$, we compute
       the expected lensing amplitude and the NFW scale-radius,
	    $2[\rho_c
       r_s]_{(a)}$ and $r_{s(a)}$, respectively, in order to ``scale''
       the amplitude of galaxy ellipticities as well as the
       radius for the $s_a$-th
	    background
       galaxy in the $a$-th cluster region: $e_{+(s_a)}\rightarrow
	    e_{+(s_a)}/[2\rho_c r_s]_{(a)}$
       and $R_{(a)s_a}\rightarrow x=R_{(a)s_a}/r_{s(a)}$. Thus, even if we
       use the same background galaxies over a
       fixed range of radii, $0.14\le R/[h^{-1}{\rm Mpc}]\le 2.8$, this
	    NFW scaling leads to different ranges of
       the scaled radii, $x$, for different clusters. The
       different amount of radial scaling requires a careful treatment
       of the radial binning, especially when comparing the lensing
       distortion profiles with and without NFW scaling. In the
following we use the different binning schemes depending on either case
	    studying the stacked lensing profile or studying the
	    scatters of 50 cluster lensing profiles relative to the NFW
	    prediction, which are summarized by the procedures (2a) and
	    (2b) below.
 \item[(2a)] {\it Stacked lensing analysis with NFW scaling
       implementation} -- As in Figure~\ref{fig:dSigma}, we will study
       the stacked distortion profile of 50 clusters after the NFW
       scaling of each cluster. Similarly to Figure~\ref{fig:dSigma}, we
	    will use the 32
       logarithmically-spaced bins in the ``scaled'' radius, $x$, where
	    we used exactly the same background
       galaxies behind the 50 clusters.  After stacking 50 clusters, we
       can expect a significant detection of the lensing signal at each
       radial bin, as implied from Figure~\ref{fig:dSigma}. However, the
	    above
       NFW scaling transforms the original radial range to a different
      range of the scaled radius $x$ for each cluster. Hence, the sample of background
	    galaxies in each bin of $R$ or $x$ radii differ from each
	    other. Nevertheless, since the stacked lensing has a sufficiently high
	    $S/N$ at each bin, we checked that the NFW scaling almost
	    conserves the total $S/N$ value (exactly speaking, it causes
	    only about 0.5\% fractional change).  We estimate
	    the representative value of each radial bin in a similar
	    manner to Eq.~(\ref{eq:r}).
 \item[(2b)] {\it Studying the scatters of lensing profiles for 50 clusters}
       -- As we discussed in \S~\ref{sec:stacked_nfw}, we monitor the
	    scatters of 50 cluster lensing profiles relative to the NFW
	    prediction in order to address the existence of the
	    universal NFW profile. To quantify the scatters, we compute
	    the $d^2$ value for either case with or without NFW scaling
	    (see Eqs.~\ref{eq:d} and \ref{eq:d_nfw} for the
	    definition). In doing this, we need to probe the ``shape''
	    of lensing profile for each cluster, and in other words each
	    radial bin needs to be in the signal dominated regime on
	    an individual cluster basis.  Hence, if we take the 32 bins as
	    in the stacked lensing analysis, each radial bin suffers
	    from the shape noise contamination due to too low number
	     density of background galaxies in each bin.  To tackle this
	     obstacle, 
	     we employ 8 logarithmically-spaced
	    bins in the range $0.14\le R/[h^{-1}{\rm Mpc}]\le 2.8$ for
	    each cluster. Figure~\ref{fig:sn_8bin} shows the expected
	    $S/N$ at each radial bin for the 50 clusters. The figure
	    shows that 319 data points among 400 points, corresponding
	    to 79\% of 400 data points, are expected to have the $S/N$
	    value greater than unity. Hence the 8 bins seem suitable for
	    our purpose. Table~\ref{tab:x-ray} gives the total
	    $S/N$ of each cluster when employing the 8 bins.  However,
	    the expected lensing signal at each radial bin would be
	    still noisy. To avoid any artifact arising from the noise
	    dominated bins, we transform each of the original bins in
	    $R$ to the corresponding bin in the scaled radius $x$ after
	    the NFW scaling, rather than redefining the radial bins for
	    a fixed range of $x$. With this binning, each radial bin
	    before and after the NFW transformation preserves the same
	    background galaxies. Hence, this binning method preserves
	    the $S/N$ value in each radial bin as well as the total
	    $S/N$ value for each cluster, before and after the NFW
	    scaling, as can be found from Eqs.~(\ref{eq:d}) and
	    (\ref{eq:d_nfw}) mathematically. As a result, the different
	    clusters cover different ranges of the scaled radius $x$.
\end{itemize}

Figure~\ref{fig:fnfw} shows the stacked lensing profiles after
implementing the NFW scaling (the above case 2a), using the halo mass
proxies based on the HSE assumption or the gas mass, respectively.  We
again note that, to have a fair comparison with Figure~\ref{fig:dSigma},
we have used exactly the same background galaxies.  The solid curve in
each panel is {\it not} a fit, but rather is the NFW prediction
(Eq.~\ref{eq:fnfw}), including the reduced shear correction
$1/[1-\kappa^{\rm NFW}(x)]$, where we used the best-fit NFW model to the
stacked distortion profile in Figure~\ref{fig:dSigma}. The reduced shear
correction is not large (by up to 20\% in the amplitude at the inner bins) over the range
of radii,
as can be
found from Figure~\ref{fig:test_nfw}.  The figure shows that the stacked
profile is in excellent agreement with the NFW prediction, to within the
errorbars. This agreement supports the existence of NFW profile in the
clusters, and implies that the X-ray inferred mass indeed gives a proxy
of the genuine mass for each cluster. To be more precise, the reduced
chi-square is $\chi^2/{\rm d.o.f}=31.3/32$ or $30.7/32$ for the HSE or
gas mass case, respectively, compared to $\chi^2/{\rm
d.o.f}=22.5/(32-2)$ in Figure~\ref{fig:dSigma}.

%
\begin{figure}
 \centering{
 \includegraphics[width=0.48\textwidth]{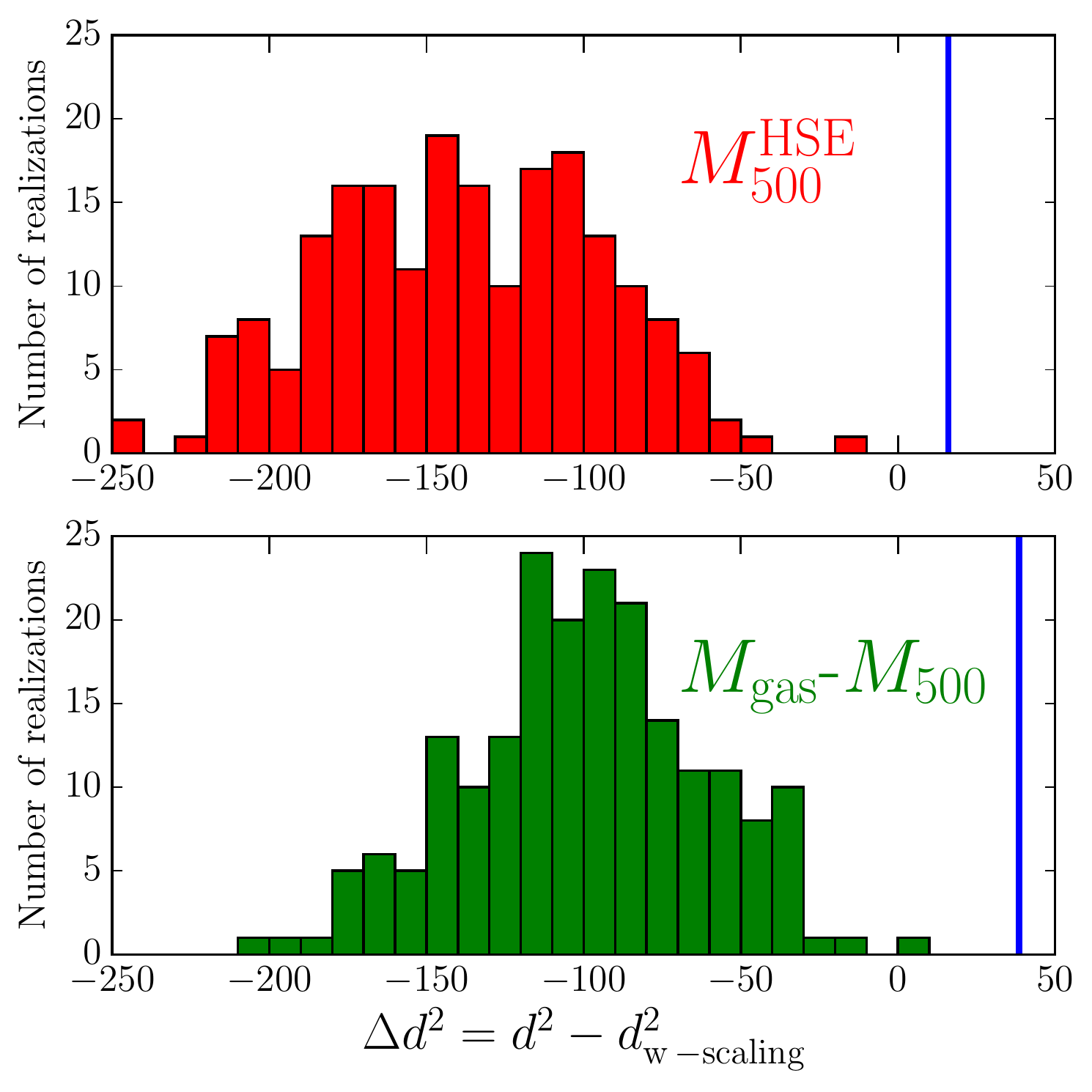} }
 \caption{A test of the performance of the NFW scaling analysis in
 Figure~\ref{fig:fnfw}. We randomly assigned the X-ray inferred mass to
 each cluster, redid the scaling analysis, and then computed the
 $d^2$ difference (Eq.~\ref{eq:dd2}).
 The histogram shows the distribution of 200 random realizations, which
 can be compared to our main result shown by the vertical line for
 either the HSE mass or the gas mass case in the upper or lower panels,
 respectively.  All the random realizations give a negative value of
 $\Delta d^2$, and any of those does not reproduce the measurement
 value.  Compared to the mean and variance of the random realizations,
 the measured $\Delta d^2$ value is away from the mean at 3.6 and
 3.7$\sigma$ for the HSE and gas mass cases, respectively.
 \label{fig:d2_Mrandom} }
\end{figure}
 \begin{figure}
  \centering{
\includegraphics[width=0.48\textwidth]{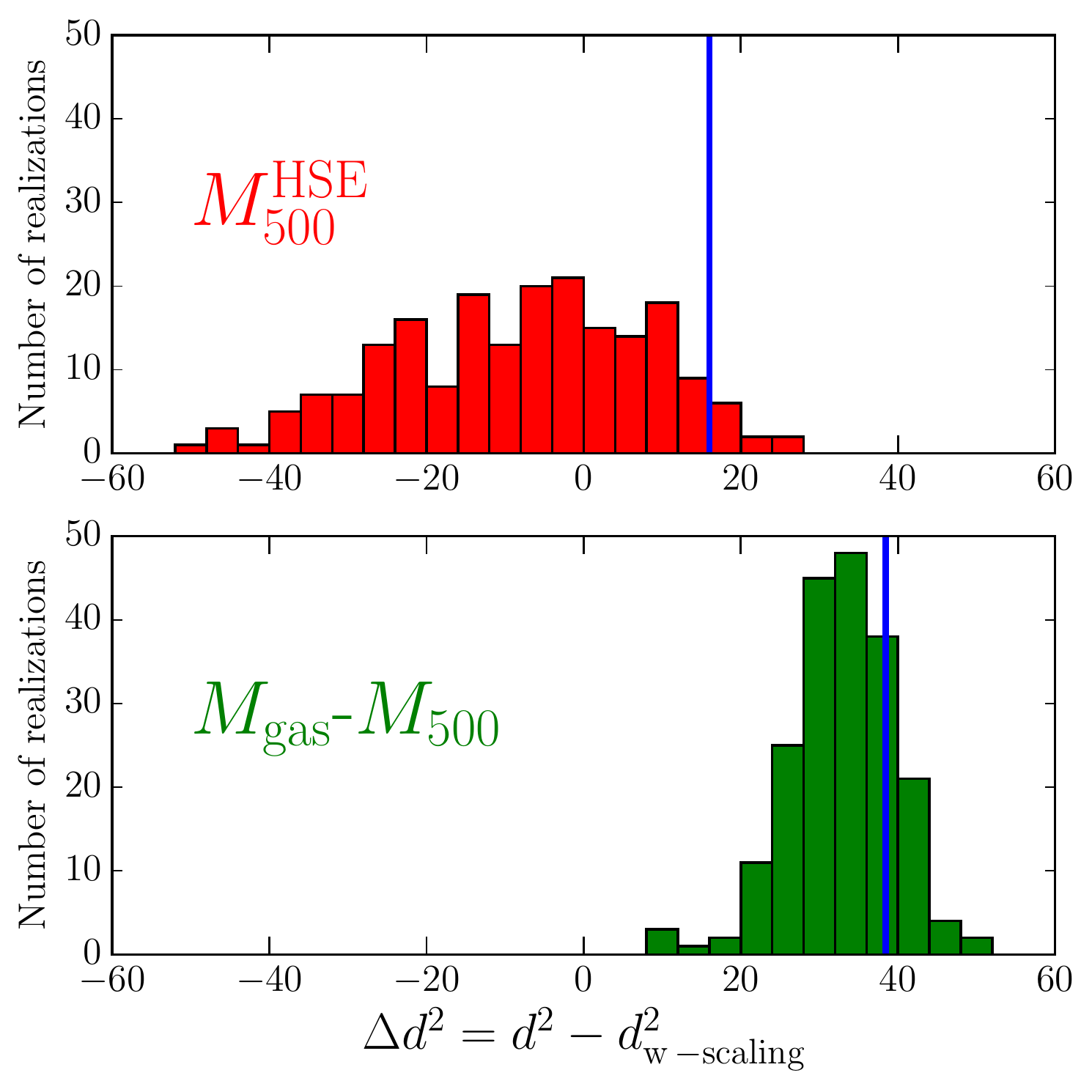}}
  \caption{As in the previous figure, but show effects of the
  statistical errors in X-ray inferred halo mass of each cluster on the
  $\Delta d^2$ value.  Here, we added a random scatter to each halo mass
  by an amount of the quoted errorbar in Table~\ref{tab:x-ray} assuming
  the Gaussian distribution, redid the weak lensing analysis with NFW
  scaling, and then computed the $\Delta d^2$ value for each
  realization.  \label{fig:d2_Merrors}}
 \end{figure}
 \begin{figure}
  \centering{
\includegraphics[scale=1.0]{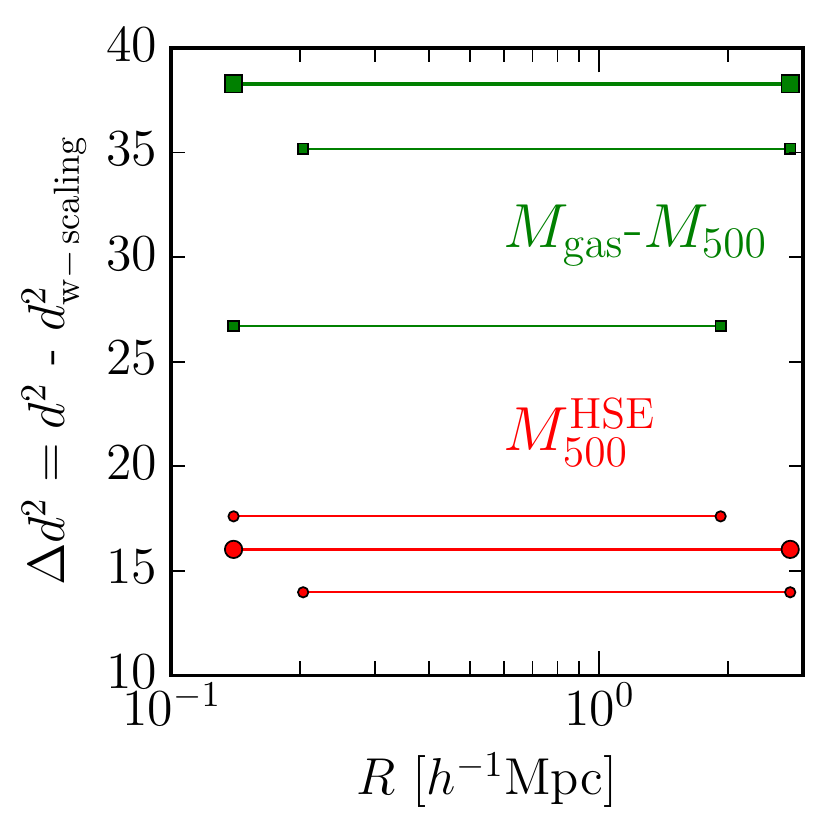}}
  \caption{The figure shows the change in $\Delta d^2$ (the vertical axis)
when using the different range of radii (the horizontal axis) used in
the analysis. The bold lines are the results for our fiducial choice:
 $0.14\le R/[h^{-1}{\rm Mpc}]\le 2.8$ and 8 logarithmically-spaced bins
for all the 50 clusters. The other lines are the results when excluding
the innermost or outermost bin from the analysis, respectively, for the
the HSE or gas mass proxy cases, respectively.
 }
\label{fig:d2_data_range}
 \end{figure}
Now we compare the scatters of 50 cluster lensing profiles with and
without implementation the NFW scaling, quantified by the $d^2$ value
(Eqs.~\ref{eq:d} and \ref{eq:d_nfw}), in order to address the existence
of {\it universal} NFW profile.  By using the above method (2b), we find
the difference between the $d^2$ values with and without NFW scaling as
%
\begin{eqnarray}
 \Delta d^2&\equiv& d^2-d^2_{\rm w-scaling}\nonumber\\
 &=&
  \left\{
\begin{array}{ll}
 543.2-527.1\simeq (4.0)^2, & \hspace{1em}\mbox{(HSE)}\\
 543.2-504.6\simeq (6.2)^2, & \hspace{1em}\mbox{($M_{\rm gas}$-$M_{500c}$)}
\end{array}
	 \right.
  \label{eq:dd2}
\end{eqnarray}
%
Thus the NFW scaling for both the HSE and gas mass cases leads to the
smaller $d^2$-values, meaning the smaller scatters of lensing profiles
relative to the NFW profile than the scatters without NFW scaling. The
smaller $d^2$ value for the gas mass implies that the gas mass gives a
better proxy of the underlying true masses of clusters (at least the
relative mass differences between different clusters).
Thus the NFW scaling gives about 4- or 6-$\sigma$ improvement for the
HSE or gas mass, respectively, assuming that the $d^2$ distribution
obeys a $\chi^2$-distribution.  We checked that, even if we use 16 bins
instead of our fiducial 8 bins, the $d^2$ values themselves get enlarged
because each bin is more in the shape noise dominated regime, but the
$d^2$ difference, the $\Delta d^2$ value, is almost unchanged.

The smaller $d^2$ value due to the NFW scaling arises from two parts:
the scaling of lensing profile amplitude (or background galaxy
ellipticities) and the scaling of cluster-centric radius. The two
scalings are specified by halo mass and concentration of each cluster:
$e_+/[2\rho_c r_s]\propto M_{500c}^{-1/3}c_{500c}^{-2}$ and
$x=R/r_s\propto M_{500c}^{-1/3}c_{500c}$, respectively. If we include
only the scaling of background galaxy ellipticities, without the radial
scaling, we found $d^2_{\rm w-scaling}=512.9$ or $526.5$ for the HSE or
gas mass case, which equivalently correspond to $\Delta d^2\simeq
(5.5)^2$ or $(4.1)^2$, respectively. That is, the HSE case shows an even
greater improvement in $\Delta d^2$ compared to Eq.~(\ref{eq:dd2}).  On
the other hand, if we include only the scaling of radius, but without
the scaling of galaxy ellipticities, $d^2=547.1$ or $522.3$, which
correspond to $\Delta d^2\simeq -3.9$ or $(4.6)^2$, respectively. Thus,
for the HSE case, the radial scaling does not appear to be adequate, and
rather gives a positive $\Delta d^2$.  For the gas mass case, both the
two scalings about equally contribute to the reduce in the $d^2$ value
or the significance of the smaller scatters. 

Figure~\ref{fig:scatters_data} shows the contribution of each cluster to
the $d^2$-value, which shows the argument of Eqs.~(\ref{eq:d}) or
(\ref{eq:d_nfw}) at each radial bin for each of 50 clusters. The total
$d^2$ value is obtained by summing the square of each curve over the 8
radial bins and 50 clusters. Table~\ref{tab:x-ray} gives the total
$d^2$-value for each cluster.  The figure shows that, although it looks
noisy, the NFW scaling reduces the scatters. One might notice some
outlier clusters: the clusters, which have top three largest $d^2_{\rm
w-scaling}$ values (see Table~\ref{tab:x-ray}), are A781, A209, and A697
for the HSE case, while A781, A2645 and A750 for the gas mass case,
respectively.  For example, the bold curve denotes the A781 cluster, which
is a more than $3\sigma$ outlier than the mean. The mass distribution of
A781 displays three prominent peaks, indicating that the cluster is in
the phase of ongoing merger \citep{Wittmanetal:14}.

For further clarification, Figure~\ref{fig:scatters_cl25} shows the
scatters for 25 clusters which are a half of 50 clusters that have the
larger value of $\Delta d^2$ (Eq.~\ref{eq:dd2}), i.e. show the better
improvement of NFW scaling analysis, than other 25 clusters (see
Table~\ref{tab:x-ray}).  The reduce in the scatters due to the NFW
scaling is more evident. In this case, the cumulative value of $\Delta
d^2$ for 25 clusters, $(\Delta d^2)^{1/2}=(86.2)^{1/2}\simeq
9.3$. Similarly, $(\Delta d^2)^{1/2}\simeq (80.6)^{1/2}\simeq 9.0$ for
the HSE mass.  The other 25 clusters yield smaller or even positive
$\Delta d^2$ values, probably due to the complex mass distribution or
the inaccuracy in X-ray inferred halo masses.

To draw a more robust conclusion, we make several tests of our
results. In Figure~\ref{fig:d2_Mrandom}, we studied how the scatters of
50 lensing profiles are enlarged if we implement the NFW scaling
analysis by randomly assign the X-ray inferred halo mass to each cluster
(without repeated use of X-ray mass).  All the 200 random realizations
have a negative value of $\Delta d^2$, and any of the random
realizations cannot reproduce a similar positive value to the measured
$\Delta d^2$ (the vertical line) for both the HSE and gas mass cases.
To be more quantitative, the measured value $\Delta d^2$ is away at 3.6
and 3.7 $\sigma$ for the two cases, respectively, compared to the mean
and variance of the random realization distribution. These results give
another support on the existence of NFW profile in the 50 clusters.

One important source of uncertainties in the method is a residual
uncertainty in the X-ray inferred halo mass or a possible effect of
intrinsic scatter in the mass scaling relation of X-ray observable.
Figure~\ref{fig:d2_Merrors} shows how the statistical errors of X-ray
inferred mass affect the $\Delta d^2$ value. To be more precise, we
added a random scatter to halo mass of each cluster
assuming the Gaussian distribution with variance given by the quoted
errorbar of each X-ray mass in Table~\ref{tab:x-ray},
i.e. $M^\prime_{500 (a)}=M_{\rm 500 (a)}^X+\delta M_{(a)}$, treated the
shifted mass as its true mass, and then redid the NFW scaling
analysis. The figure shows that adding the random scatter to each
cluster tends to decrease $\Delta d^2$, implying that the central value
of the X-ray inferred mass is indeed closer to the underlying true
mass. The distribution of $\Delta d^2$ is wider for the HSE mass, but
this would be ascribed to the larger errors of HSE mass than those for
the gas mass as can be found from Figure~\ref{fig:m-z}. Again
encouragingly, even if adding the random errors to the gas mass, the
resulting $\Delta d^2$ values are positive, supporting that the gas mass
is a better proxy of the genuine cluster mass as in
Figure~\ref{fig:fnfw}.

Although we have used the fixed range of the original comoving radius,
$0.14\le R/[h^{-1}{\rm Mpc}]<2.8$, for all the clusters as our fiducial
choice, Figure~\ref{fig:d2_data_range} shows how the results are changed
if excluding the inner- or outermost radial bin of 8
logarithmically-spaced bins from the analysis. The figure shows that,
for the X-ray gas mass proxy, excluding the outer- or innermost bin
degrades the NFW scaling or reduces the $\Delta d^2$ values, suggesting
that the wider range of radii is important to capture the curvature of
the mass profile.  On the other hand, for the HSE mass case, excluding
the outermost bin increases the $\Delta d^2$, again implying that the
HSE mass estimate might not be as accurate to infer the genuine mass as
the gas mass and involve residual systematic errors.

\subsection{Discussion and Implications}

\subsubsection{Comparison with $N$-body simulations}
 \begin{figure}
  \centering{
\includegraphics[width=0.5\textwidth]{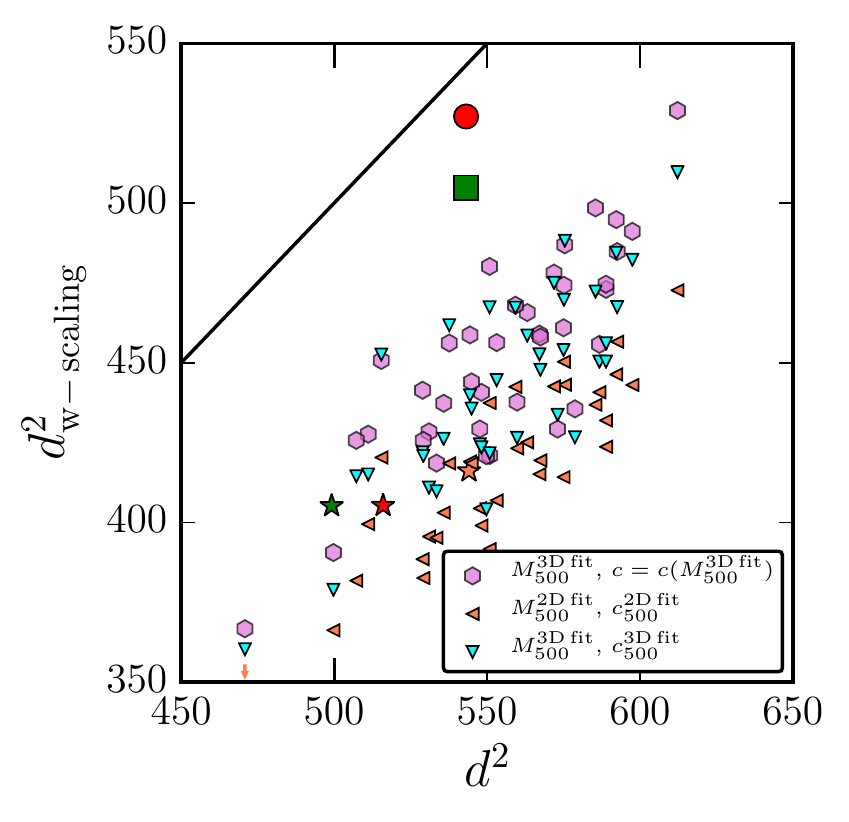}}
  \caption{Comparison of the measurements and the simulation results for
  the $d^2$-values (Eqs.~\ref{eq:d} and \ref{eq:d_nfw} for their
  definitions).  The bigger-size circle and square symbols are the
  measurement results for the HSE and gas mass cases, respectively. The
  other symbols are the simulation results as in
  Figure~\ref{fig:test_nbody}, but including the effect of intrinsic
  ellipticities of background galaxies that are taken from actual Subaru
  data of each cluster region (see text for details).
  The tilted triangle and hexagon symbols are the simulation results
  when using the best-fit NFW parameters of 3D mass profile for each
  halo, the NFW parameters of 2D projected lensing profile or the
  best-fit mass of 3D mass profile, but the concentration inferred from the
  scaling relation $c_\Delta=c_\Delta(M_\Delta;z)$, respectively (see
  Figure~\ref{fig:test_nbody} for details). The third case is intended
  to mimic what we did for the actual measurements. We show the
  simulation results for 40 realizations of background galaxy
  ellipticities. Note that, for each realization, we computed the three
  simulation results; each of the triangle and hexagon symbols has the
  same $d^2$ value in the horizontal axis, but different $d^2_{\rm
  w-scaling}$ values in the vertical axis.  For comparison, the
  orange-color star symbol denotes one particular realization that has a
  similar $d^2$ value to the measurement for no NFW scaling case (the
  horizontal axis). The two star symbols in the left-lower corner are the
  results when using the same realization of background galaxies as in
  the orange-color star symbol, but using the analytical NFW profiles
  for the $d^2$ calculations. Note that the arrow in the lower-left
  corner denotes the simulation result that is below the plotted range.
  \label{fig:d2-vs-nbody}}
 \end{figure}
Is the NFW scaling results in Figures~\ref{fig:scatters_data} and
Eq.~(\ref{eq:dd2}) as expected from a viewpoint of $\Lambda$CDM
structure formation model?  To address this question,
Figure~\ref{fig:d2-vs-nbody} compares the measurement results and the
$N$-body simulated halos, as in Figure~\ref{fig:test_nbody}, in a
two-dimensional space of the $d^2$ values with and without NFW
scaling. To make a fair comparison, we included the effect of intrinsic
galaxy ellipticities on the simulation results. To be more precise, (1)
we first populated, into each region of simulated halos, the background
galaxies taken from the corresponding Subaru cluster data (matched in
descending order of halo masses), (2) made a random rotation of
orientation of each galaxy ellipticity, which erases the coherent
lensing signal of each Subaru cluster, (3) simulated the
``observed'' galaxy ellipticity of each background galaxy by adding both
the lensing distortion of simulated halo and the intrinsic shape, (4)
performed the hypothetical lensing measurements with and without NFW
scaling, and (5) computed their $d^2$ values, respectively.
To account for the statistical variance of intrinsic ellipticities, we
generated 40 realizations of the $N$-body simulation results: we redid
the $d^2$ calculations after random rotation of background galaxies.
For the simulation results, we consider the three cases similarly to
Figure~\ref{fig:test_nbody}: the lensing analysis with NFW scaling when
using the best-fit NFW parameters of 3D mass profile, the NFW parameters
of 2D distortion profile, or the best-fit halo mass of 3D profile, but
using the halo concentration inferred from the $c$-$M$ relation,
respectively. The third case is closest to what we did for the actual
data.  First of all, the simulation results without NFW scaling, denoted
by the $d^2$ values in the horizontal axis, fairly well reproduce the
measurements on average, reflecting that the statistical errors in the
$d^2$ value are dominated by the shape noise. Also note that the
horizontal spread of the simulation realizations is roughly given by
$\sqrt{d^2}\simeq \sqrt{550}\simeq 23$.  However, all the simulation
results with NFW scaling, $d^2_{\rm w-scaling}$ in the vertical axis,
are systematically smaller than the measured values.  Thus this
disagreement suggests that we do not properly consider some effects
inherent in the measurements on the simulation results.  For comparison,
the star symbols show the results when using analytical NFW halos to
compute the $d^2$ values where we used the X-ray proxy masses for
the HSE or the gas mass to compute the NFW lensing profile of each
cluster.  The difference between the analytical NFW halos and the
simulation results is due to the complexity of mass distribution in the
simulated halos, such as asphericity, substructures, and the scatters of
halo concentration.

\begin{figure*}
  \centering{
\includegraphics[width=0.95\textwidth]{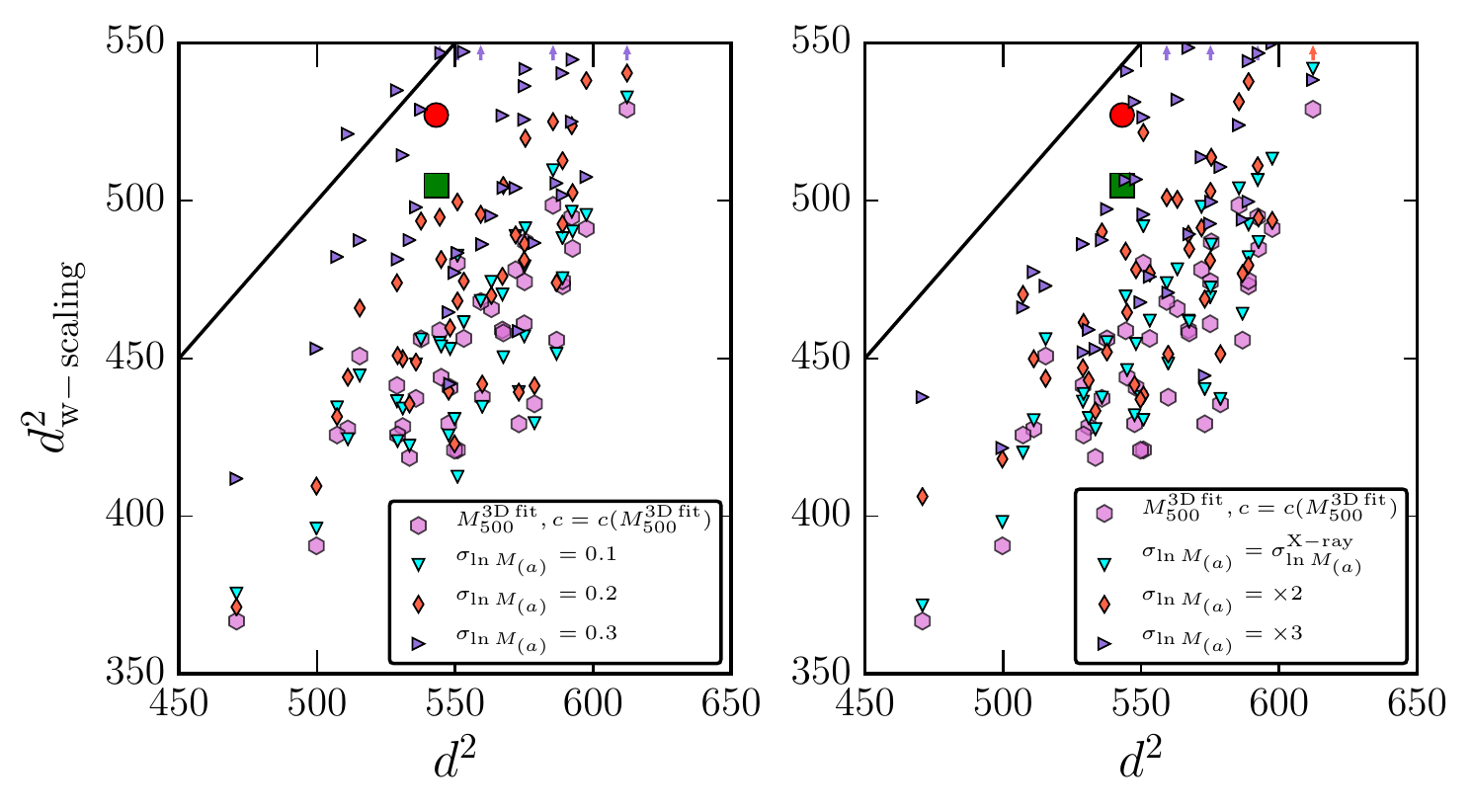}}
  \caption{Similar to the previous figure, but here included the effects
 of mass scatter of each halo on the simulation results. The hexagon
 symbols are the same as in Figure~\ref{fig:d2-vs-nbody}. For each
 realization of background galaxies, we added a random mass scatter to
 each halo, simulated the lensing analysis with NFW scaling by treating
 the shifted mass as the true mass, and then computed the $d^2_{\rm
 w-scaling}$ value. Adding the halo mass scatters tends to degrade the
 NFW scaling results or preferentially causes an up-scatter of each
 simulation result in this two-dimensional space. {\it Left panel}: The
 simulation results when adding the Gaussian mass scatters by the
 fractional errors of $\sigma_{\ln M}=0.1, 0.2$ or 0.3,
 respectively. The arrows in the upper horizontal axis denote the case
 that the simulation results are outside the range shown in this plot.
 {\it Right panel:} Similar to the left panel, but the results when
 adding a random mass scatter to each simulated cluster assuming the
 fractional error proportional to the quoted error of the gas mass proxy
 relation (assigned to the simulated halo); $\sigma_{\ln M_{(a)}}\equiv
 \sigma^X_{M_{(a)}}/M_{(a)}^X$ or a factor 2 or 3 bigger
 one. \label{fig:d2-vs-nbody_Merrors}}
\end{figure*}
A possible source to reconcile the difference between the measurements
and the simulation results in Figure~\ref{fig:d2-vs-nbody} is an
additional error or intrinsic scatter in the X-ray inferred halo mass
\citep{Staneketal:10,Okabeetal:10b}.
Figure~\ref{fig:d2-vs-nbody_Merrors} addresses this question. The left
panel shows how adding a scatter to each mass of simulated halos,
parametrized by the fractional variance $\sigma_{\ln M}=\sigma(M)/M=0.1,
0.2$ or 0.3, degrades the $d_{\rm w-scaling}^2$ values for the NFW
scaling analysis. More precisely, we randomly generated a mass scatter
$\delta M$ for each halo assuming the Gaussian distribution with
variance $\sigma_{\ln M}$, added the scatter to each halo mass as given
by $M^\prime_{(a)}=M^{\rm{2D~ fit}}_{500(a)}+\delta M$, and then
computed the $d^2_{\rm w-scaling}$ value by treating the shifted mass
$M'$ as the true mass of each simulated halo. For the sake of
comparison, we used the same 40 realizations of background galaxies as
in Figure~\ref{fig:d2-vs-nbody}, and therefore the degradation is solely
due to the mass scatters. Note that, for each realization of background
galaxies, adding the halo mass scatters changes only the $d^2_{\rm
w-scaling}$ value in the vertical axis.
The figure shows that the halo mass scatters generally degrades the NFW
scaling result or equivalently enlarge the $d^2_{\rm w-scaling}$
value. However, only the additional errors of $\sigma_{\ln M}\sim
0.2$--$0.3$ can reproduce the measurement result for the gas mass
proxy.
This might imply that
the X-ray halo mass involves an unknown, systematic error or intrinsic
scatter.

\begin{figure}
\centering{
 \includegraphics[width=0.48\textwidth]{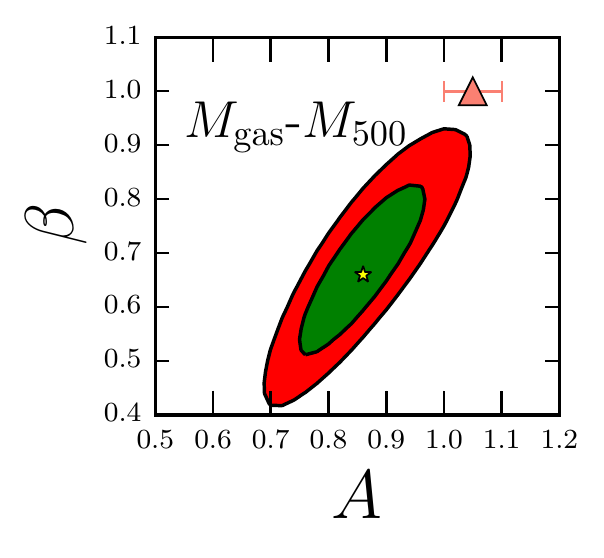}}
 \caption{Effects of variations in the halo mass proxy relation of X-ray
 gas mass on the NFW scaling results.  We model the variations as
 $M_{500c}/[10^{14}M_\odot]=A\times 11.6\times (M_{\rm
 gas}/10^{14}h^{-3/2}M_\odot)^{\beta}$, where $A=1$ and $\beta=1$ are
 our fiducial model corresponding to the self-similar scaling model. We
 estimated the best-fit parameters (the star symbol) by
 minimizing the $d^2$ value with varying the normalization and mass
 slope parameters. The two contours correspond to the regions satisfying
 the conditions $\Delta d^2=d_{\rm w-scaling}^2(A,\beta) - d_{\rm
 w-scaling}^2(A^{\rm best-fit},\beta^{\rm best-fit})= 2.3$ or 6.17,
 respectively. The triangle symbol with errorbar denotes the result when
 varying the normalization parameter $A$ alone, with fixing $\beta=1$.
}
\label{fig:Mgas-M500_three}
\end{figure}
As an alternative test, the right panel of
Figure~\ref{fig:d2-vs-nbody_Merrors} shows the effects of the quoted
errorbars in the X-ray inferred halo masses. Here we added a random mass
scatter to each cluster,
$M^{\prime}_{(a)}=M_{500(a)}^{\rm{2D~fit}}(1+\delta \ln M)$ by taking
the fractional mass error, $\sigma(M^X_{(a)})/M^X_{(a)}$, for each
cluster (see Table~\ref{tab:x-ray}) assuming the Gaussian distribution,
and then computed the $d^2_{\rm w-scaling}$ value for each realization.
Here we used the mass errors for the gas mass proxy in
Table~\ref{tab:x-ray}. Note that the mean fractional error of 50
clusters is about 0.11, but here we included variations in the errors
for different clusters.  The figure shows that, if each cluster has a
factor 2--3 larger mass error than the quoted error, the simulation
results appear to reproduce the measurements.
The mass errors of X-ray observables might underestimate the genuine
mass uncertainty, perhaps due to the limitation of the X-ray based
method or due to an unknown intrinsic scatter in the X-ray observable
and halo mass relation \citep[see also][for the similar
discussion]{Okabeetal:10b}.

\subsection{The halo mass proxy relation of X-ray observables}
\label{ssec:xraymass}
The method we have so far developed involves several assumptions. For
instance, to implement the lensing stacking with NFW scaling, we need to
assume several scaling relations: the halo mass proxy relation of X-ray
observables and the halo mass and concentration relation. In the
following we address how possible variations in these scaling relations
affect the NFW scaling results.

First we study a possible bias in the X-ray inferred halo mass. Since
the halo mass proxy relation of X-ray gas mass showed a better
performance in the NFW scaling analysis (see Eq.~\ref{eq:dd2}), we here
consider effects of possible variations in the gas mass relation on the
results. To address this, we modify Eq.~(\ref{eq:M500-Mgas}) to
parametrize the halo mass proxy relation as
\begin{eqnarray}
&& \frac{M^X_{\rm 500}}{10^{14}M_\odot} =A\times 11.6\times \left(
\frac{M_{\rm  gas}}{10^{14} h^{-3/2}M_\odot}	\right)^\beta,
\end{eqnarray}
where $A$ is the normalization parameter and $\beta$ is a slope
parameter of the halo mass dependence. The model with $A=1$ and
$\beta=1$ corresponds to the self-similar model given by
Eq.~(\ref{eq:M500-Mgas}).  With varying the two parameters, $A$ and
$\beta$, simultaneously, we estimated the halo mass of each
cluster from the X-ray gas mass based on the above proxy relation, and
then redid the NFW scaling analysis.
Figure~\ref{fig:Mgas-M500_three} shows the constraint regions
in the two parameter space.
The best-fit parameters $A=0.86\pm 0.06$ and $\beta=0.66\pm 0.10$, which
has $d^2_{\rm w-scaling}=493.8$ compared to $d_{\rm min}^2=504.6$ for
the fiducial model as given by Eq.~(\ref{eq:dd2}), corresponding to
about 3$\sigma$ improvement. Here we quoted the errorbars from the range
$\Delta d^2\le 1$ with varying both $A$ and $\beta$, although the
degeneracy between the two parameters is significant.  Thus the scatters
of 50 lensing distortion profiles prefer a weaker halo mass dependence
than predicted by the self-similar scaling relation at a 3$\sigma$
level.  Note that, if the slope parameter is fixed to $\beta=1$,
i.e. the self-scaling relation, we obtained $A=1.05\pm 0.05$ ($d^2_{\rm
min}=503.6$, almost no change from the fiducical model). These results
might be due to some residual uncertainty in our method, and would be
worth further exploring by using a larger sample of clusters or an
independent mass proxy relation such as the Sunyaev-Zel'dovich effect.

\subsection{The halo mass and concentration relation}
\label{sec:m-c}
\begin{figure*}[th]
\centering{
\includegraphics[width=0.8\textwidth]{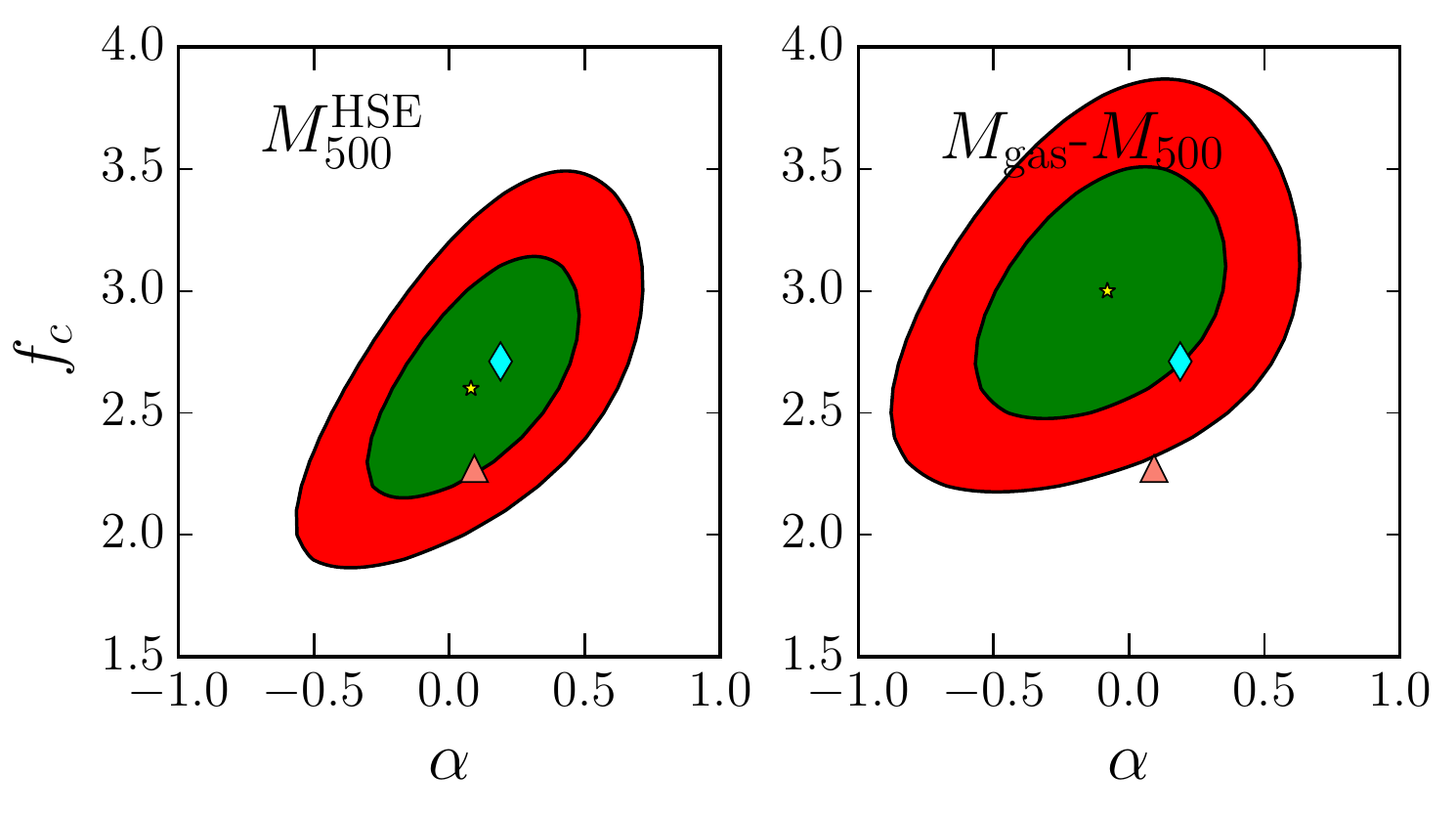}}
 \caption{Effects of variations in the halo mass and concentration relation on the
NFW scaling results, for the HSE and gas mass cases, respectively. Here
we parametrized the variations as $c(M)\propto M^{\alpha}$
(Eq.~\ref{eq:c-m_vary}), and then minimized the $d^2$-value with varying
the normalization and mass slope parameters. The star symbol denotes the
best-fit model.  The two contours correspond to the regions satisfying
the conditions $\Delta d^2=d_{\rm w-scaling}^2(A,\beta) - d_{\rm
w-scaling}^2(A^{\rm best-fit},\beta^{\rm best-fit})= 2.3$ or 6.17,
respectively.  The diamond symbol in each panel shows the parameters for
our fiducial model DK15 at the mean redshift of clusters, $z=0.23$,
while the triangle symbol denotes the parameters of
\citet{Duffyetal:08}.
}
\label{fig:m-c}
\end{figure*}
Another important model ingredient in our analysis is the scaling
relation of halo concentration with halo mass. We have so far employed
the scaling relation in DK15 as for our default model. However, other
works have proposed a different scaling relation from DK15.  For
example, \citet{Duffyetal:08} proposed a different fitting formula of
the $c$-$M$ relation, and predicts a 20--30\% lower concentration than
in DK15 for cluster-scale halos: more exactly $c_{200c}\sim 3$ or 4 for
\citet{Duffyetal:08} or DK15 for cluster-scale halos of several times
$10^{14}h^{-1}M_\odot$ masses at $z\sim 0$ (see Figure~9 in DK15).
However, we found that, even if we use the scaling relation in
\citet{Duffyetal:08} instead of DK15, it almost unchanges the $d^2_{\rm
w-scaling}$ value; more exactly, it enlarges the $d^2_{\rm w-scaling}$
value only by $\Delta d^2\simeq 1$--$2$ for the HSE and gas mass
proxy relations. Hence the current data cannot discriminate these
different models of $c$-$M$ relation.

Nevertheless one might ask whether or not our method allows us to
constrain the underlying $c$-$M$ relation.
Assuming the parametrized form of $c$-$M$ scaling relation given by
\begin{equation}
 c_{500c}(M_{500c};z)=f_c\left(\frac{M_{500c}}{4\times
		      10^{14}~h^{-1}M_\odot}\right)^{-\alpha}\times
 (1+z)^{-0.51},
 \label{eq:c-m_vary}
\end{equation}
we minimized the $d^2_{\rm w-scaling}$ value with varying the
normalization parameter and the mass slope parameter, $f_c$ and
$\alpha$. Here we took the halo mass inferred from the stacked lensing in
Figure~\ref{fig:dSigma}, $M=4\times 10^{14}~h^{-1}M_\odot$, for the
pivot mass scale, and the redshift dependence is taken from
\citet{Duffyetal:08}\footnote{Exactly speaking the fitting formula of
\citet{Duffyetal:08} gives the $c$-$M$ relation for $M_{200c}$, so we
converted the scaling relation to the relation between $M_{500c}$ and
$c_{500c}$, and found that the redshift dependence is slightly modified
from the original dependence $(1+z)^{-0.47}$ by this conversion.}.  Note
that
we fixed the mass normalization parameter to $A=1$ for the halo mass
proxy relation of X-ray observables. Figure~\ref{fig:m-c} shows that constraints
on the two parameters are significantly degenerate: the $d^2_{\rm
w-scaling}$ for the best-fit model is 526.3 or 503.5 for the HSE and gas
mass, respectively, which is slightly smaller than our fiducial model,
DK15, as found from Eq.~(\ref{eq:dd2}).  The best-fit parameters are
$f_c=2.6_{-0.2}^{+0.3}$ and $\alpha=0.08^{0.13}_{-0.12}$ for HSE, while
$f_c=3.0\pm 0.3$ and $\alpha=-0.08\pm 0.18$ for the gas mass scaling
relation.  Thus the current data prefers the amplitude of concentration
to be $c_{500c}\simeq 2.6$--3.0 for the 50 clusters of these mass
scales, which is consistent with both the theory predictions in
\citet{Duffyetal:08} and DK15 within the errorbars, but cannot well
constrain the mass slope due to the limited statistics or a narrow
range of halo masses.

\subsection{The halo mass profile}

The lensing analysis of NFW scaling rests on the assumption that the
mass distribution in clusters follows the {\it universal} NFW
profile. However, the NFW profile is the simplified prediction of
$N$-body simulations, and a further improvement in our method might be
available by employing a better model of the mass profile.

Several works have pointed out variations in the inner region of the
mass profiles. For instance, there might be variations in the inner
slope of the mass profile \citep[e.g., see][and references
therein]{Navarroetal:04}. The baryonic processes would generally affect
the inner structures, which tend to cause a greater mass concentration
in the inner region and generally breaks the universality of the total
mass profile. However, in this study, we looked into the cluster lensing
signals down to $R\simeq 0.14~h^{-1}{\rm Mpc}$ and above, and these
effects would be unlikely to largely change our results.

Another interesting effect is a possible variation in the outer mass
distribution at radii near to the virial radius or greater, as proposed
in \citet{DiemerKravtsov:14b} and \citet{Adhikarietal:14}. These works
claimed that the logarithmic slope of massive halos steepens more
sharply than the NFW predicts, at the outer regions $R\simgt
0.5R_{200c}$, depending on the details of mass accretion and assembly
history.
This breaks to some extent the universality of NFW profile at
these outer radii.
Note that our analysis uses the lensing profile up to
$R_{\rm max}=2.8~$Mpc/$h$, which corresponds to $R_{\rm max}\sim
R_{200c}$ for the 50 clusters.
We tested the prediction of \citet{DiemerKravtsov:14b}
by using the fitting
formula for a typical accretion history that is kindly made available to
us by Surhud More \citep{Moreetal:15b}. However, we found that the current datasets cannot
discriminate the steepened profile and the NFW profile at the outer
radii. This would be interesting to further explore with an enlarged
sample of clusters.

\subsection{A residual bias in source redshift}

As we discussed above, our results imply that the X-ray inferred mass
may systematically underestimate the true mass: we found a possible bias
of 5\% level, although it is not significant (at a $1\sigma$
level).  Eq.~(\ref{eq:nfwlens_cM}) suggests that a 10\% bias in halo
mass corresponds to about 3\% bias in the lensing amplitudes. This is a
tiny amount, and may imply a residual error in the source redshift
estimation.
Due to the limited color information of
the current data (mostly only 2 colors), we cannot resolve this, but a
further study is definitely worth exploring. For the same reason, it is
worth further looking into a possible remaining systematic error in the
shape measurement.

\section{Conclusion and Discussion}
\label{sec:conclusion}

In this paper, we have developed a novel method of measuring the cluster
lensing distortion profiles along the NFW prediction, one of the most
important predictions of CDM structure formation model. The method
measures the cluster lensing profiles by stacking the ``scaled''
amplitudes of background galaxy ellipticities as a function of the
``scaled'' centric radius according to the NFW prediction of each
cluster. To apply this method to real data, we combined the independent
datasets for a nearly mass-selected sample of 50 massive clusters that are the
Subaru weak lensing catalog in \citet{Okabeetal:13} and their X-ray
observables of {\it XMM} and/or {\it Chandra} satellites in
\citet{Martinoetal:14}. Here we used the X-ray observables to infer the
NFW parameters of each cluster; more precisely, we used the halo mass of
each cluster based on the halo mass proxy relation of X-ray observables,
either the hydrostatic static equilibrium or the self-similar scaling
relation of gas mass, and inferred the halo concentration from the
$c$-$M$ relation found in N-body simulations of DK15.
We found a 4 -- 6$\sigma$ level evidence of the existence of universal
NFW profile in the 50 massive clusters (see Eq.~\ref{eq:dd2} and
Figures~\ref{fig:fnfw} -- \ref{fig:d2_Mrandom}). To derive these results
we have carefully studied a proper radial binning of the lensing
distortion measurement and how to define the representative central
value of each radial bin taking into account the cluster-centric
distances and the lensing weights of background galaxies in the annulus.
Our results give a proof of concept of the method we developed in this
paper.

However, the improvement in the scatters of 50 cluster distortion
profiles due to the NFW scaling analysis is not as much as expected from
theory using simulations of cluster based on high-resolution $N$-body
simulations (Figures~\ref{fig:d2-vs-nbody} and
\ref{fig:d2-vs-nbody_Merrors}). We discussed that, in order to
reconcile the difference between the measurements and the simulation
expectation, we need to introduce additional halo mass scatters to each
cluster, by an amount of $\sigma(M)/M\sim 0.2$--$0.3$ (see
Figure~\ref{fig:d2-vs-nbody_Merrors}). This implies intrinsic scatters
in the halo mass and X-ray observable relation \citep{Okabeetal:10b}.
We also argued that the discrepancy might be due to an imperfect halo
mass proxy relation of the X-ray observables (see
\S~\ref{ssec:xraymass}). Hence it would be worth further exploring the
method by combining different observables of clusters. A promising
example is the Sunyaev-Zel'dovich (SZ) effect. By using or combining the
X-ray, optical richness and SZ effects to develop a well-calibrated
relation between halo mass and cluster observables for a suitable sample
of massive clusters, we can explore a further improvement in
constraining the universality of cluster mass distribution. In addition,
we throughout used the model $c$-$M$ relation to infer the halo
concentration of each cluster. In other words, we ignored intrinsic
scatters of halo concentration that is known to exist even for halos of
a fixed mass scale from simulation based studies. If we can use
observables to estimate halo concentration for each cluster, it might
improve the NFW scaling results. For example, the concentration of
member galaxies might be a good proxy of halo concentration on
individual cluster basis. This would be worth exploring.

Our method offers various applications. First, we inversely use the weak
lensing analysis of NFW scaling to infer the underlying true relation
between halo mass and cluster observables.
We made the initial attempt
of this possibility in \S~\ref{ssec:xraymass}.
Since the NFW scaling method up- or down-weights less or more massive
clusters in order to make their profiles to be in the similar
amplitudes, it can be applied to halos over a wider range of mass scales
as long as the clusters in the sample follow the universal NFW
profile. In this sense this method would be less sensitive to the
selection effect of clusters in a sample.  Secondly, we can use this
method to explore the underlying true form of the halo mass profile or
the halo mass scaling relation with observables, as we attempted in
\S~\ref{sec:m-c}.  As claimed in \citet{DiemerKravtsov:14b}, massive
clusters might display a steeper profile at the outer radii around or
beyond the virial radius than predicted by NFW model, depending on the
mass accretion history. By subdividing clusters into subsamples using a
proxy to infer the mass accretion history, e.g. high or low
concentration, we can use the NFW scaling analysis to explore the
deviations from NFW prediction at the outer radii. This is a direct test
of the hierarchical CDM structure formation model, and will be very
interesting to explore.

The weak lensing measurements of 50 massive clusters we used in this
paper seem to be still limited by statistics, mainly due to a low number
density of background galaxies, which we needed to take in order to
define a secure sample of background galaxies based on two passband data
alone. Hence our method can be further improved by increasing background
galaxies, based on photo-$z$ information of more color information
\citep{Medezinskietal:13}. We can also combine the lensing
magnification bias measurement to improve the statistics. On-going
wide-area optical surveys such as the HSC survey and the DES survey
promise to provide us with a much larger, well-calibrated sample of
massive clusters, so it would be interesting to apply the method
developed in this paper to those datasets in combination with other
wavelength surveys such as X-ray or SZ effects.


\section*{Acknowledgments}

We thank Yasushi Suto for useful discussion which initiated the idea of
this work, and also thank Eiichiro Komatsu, Surhud More and Masamune
Oguri for useful discussion. We also thank the LoCuSS collaboration for
allowing us to use the published results of weak lensing measurements
and X-ray observables in our study.  We thank Benedikt Diemer for making
their code to compute the halo mass and concentration relation publicly
available to us.  MT and NO are supported by World Premier International
Research Center Initiative (WPI Initiative), MEXT, Japan, by the FIRST
program ``Subaru Measurements of Images and Redshifts (SuMIRe)'', CSTP,
Japan. RM was supported by the Department of Energy Early Career Award
program.  MT is supported by Grant-in-Aid for Scientific Research from
the JSPS Promotion of Science (No.~23340061 and 26610058), MEXT
Grant-in-Aid for Scientific Research on Innovative Areas “Why does the
Universe accelerate? - Exhaustive study and challenge for the future -”
(No. 15H05893), and by JSPS Program for Advancing Strategic
International Networks to Accelerate the Circulation of Talented
Researchers.  MT was also supported in part by the National Science
Foundation under Grant No. PHYS-1066293 and the hospitality of the Aspen
Center for Physics.  NO and RT are supported by Grant-in-Aid for
Scientific Research from the JSPS Promotion of Science (No. 26800097 and
No. 25287062), respectively.  NO is also supported by the Funds for the
Development of Human Resources in Science and Technology under MEXT,
Japan.  RT is also supported by Hirosaki University Grant for
Exploratory Research by Young Scientists.  Numerical computations were
carried out on Cray XT4 at Center for Computational Astrophysics, CfCA,
of National Astronomical Observatory of Japan.

\bibliographystyle{apj}
\bibliography{refs}

\end{document}